\newif\ifpreprint
\preprinttrue

\ifpreprint
  \documentclass[11pt, 
                 numbers=noenddot,
                 headings=normal, 
                 DIV16, BCOR10mm]{scrartcl}
  \usepackage{fancyhdr}
  \usepackage{natbib}
  \usepackage{amsmath,amssymb}
  \usepackage{mathtools}
  \usepackage{authblk}
  \usepackage[hyphens]{url}
  \usepackage[linktocpage,breaklinks]{hyperref}
  \usepackage{color}
  \definecolor{comment}{rgb}{0.0,0.6,0.0}
  \usepackage{listings}
  \lstdefinestyle{cppstyle}{language=C++, basicstyle=\ttfamily\small, 
    extendedchars=true, escapeinside={/*@}{@*/},
    breaklines=true, breakatwhitespace=true,
    numbers=left, numberstyle=\tiny,
    xleftmargin=5pt,
    keywordstyle=\color{blue}\bfseries,
    stringstyle=\color{red}\ttfamily,
    commentstyle=\color{comment}\ttfamily,
    morecomment=[l][\color{magenta}]{\#}
  }
  \lstnewenvironment{c++}[1][]{\lstset{style=cppstyle}}{}
  \newcommand{\cpp}{\lstinline}
\else
  \documentclass{ansarticle}
\fi

\usepackage[utf8]{inputenc}
\usepackage{overpic}
\usepackage{xspace}
\usepackage{subcaption}

\newcommand{\abs}[1]{{\lvert#1\rvert}}

\newcommand{\foamgrid}{\textsc{FoamGrid}\xspace}
\newcommand{\gmsh}{\textsc{gmsh}\xspace}
\newcommand{\paraview}{\textsc{ParaView}\xspace}
\newcommand{\uggrid}{\textsc{UGGrid}\xspace}
\newcommand{\yaspgrid}{\textsc{YaspGrid}\xspace}
\newcommand{\geometrygrid}{\textsc{GeometryGrid}\xspace}
\newcommand{\dune}{\textsc{Dune}\xspace}
\newcommand{\dumux}{\texorpdfstring{Du\-Mu$^\text{x}$\xspace}{DuMuX\xspace}}
\newcommand{\dunemodule}[1]{\texttt{#1}}
\newcommand{\file}[1]{\texttt{#1}}

\newtheorem{interfacerule}{Rule}

\title{The Dune FoamGrid implementation\\ for surface and network grids}
\author[1]{Oliver Sander}
\author[2]{Timo Koch}
\author[2]{Natalie Schröder}
\author[2]{Bernd Flemisch}
\affil[1]{{\normalsize TU Dresden, Institute for Numerical Mathematics, oliver.sander@tu-dresden.de}}
\affil[2]{{\normalsize University of Stuttgart, Institute for Modelling Hydraulic and Environmental Systems, \{timo.koch, natalie.schroeder, bernd.flemisch\}@iws.uni-stuttgart.de}}
\ifpreprint
\else
  \runningtitle{\foamgrid}
  \runningauthor{O.~Sander, T.~Koch, N.~Schröder, B.~Flemisch}
\fi

\begin{document}

\maketitle

\begin{abstract}
We present \foamgrid, a new implementation of the \dune grid interface.  \foamgrid implements
one- and two-dimensional grids in a physical space of arbitrary dimension, which allows for grids for curved domains.
Even more, the grids are not expected to have a manifold structure, i.e., more than two elements can share a common
facet.  This makes \foamgrid the grid data structure of choice for simulating structures such as foams,
discrete fracture networks, or network flow problems. \foamgrid implements adaptive non-conforming refinement
with element parametrizations.  As an additional feature it allows removal and addition of elements in an
existing grid, which makes \foamgrid suitable for network growth problems.
We show how to use \foamgrid, with particular attention to the extensions of the grid interface needed to handle non-manifold
topology and grid growth.  Three numerical examples demonstrate the possibilities offered by \foamgrid.
\end{abstract}

\section{Introduction}

Various simulation problems are posed on domains that are not open subsets of a Euclidean space.  Frequently, such domains
are surfaces or curves embedded in a higher-dimensional Euclidean space.  Equations on such domains, sometimes called geometric
partial differential equations, comprise diffusion and transport on the surface~\citep{dziuk_elliott:2007b},
flow problems~\citep{nitschke_voigt_wensch:2012,reuther_voigt:2015},
and phase-field equations~\citep{witkowski_backofen_voigt:2012}.
Sometimes, movement of the surface itself is modeled~\citep{dziuk_elliott:2007},
and this movement may couple with processes on the surface~\citep{gross_reusken:2011}.

\begin{figure}
 \begin{center}
 \includegraphics[width=\textwidth]{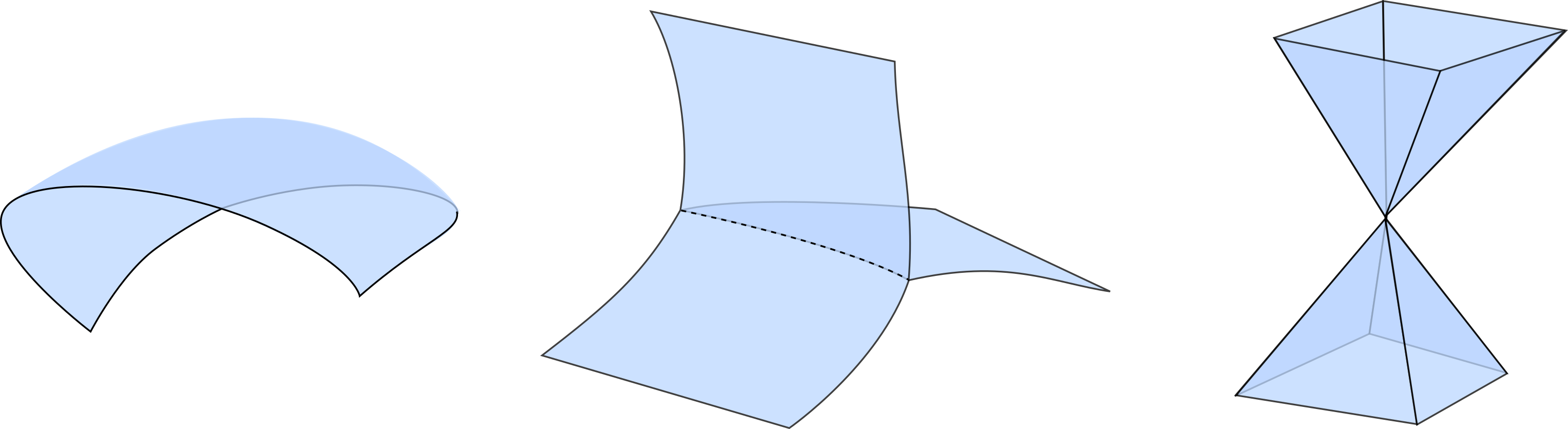}
 \end{center}
 \caption{Computational domains might not be topological manifolds.  Left: manifold; center: T-junction;
  right: touching point}
 \label{fig:non_manifold_sets}
\end{figure}

\begin{figure}
 \begin{center}
  \includegraphics[height=0.24\textheight]{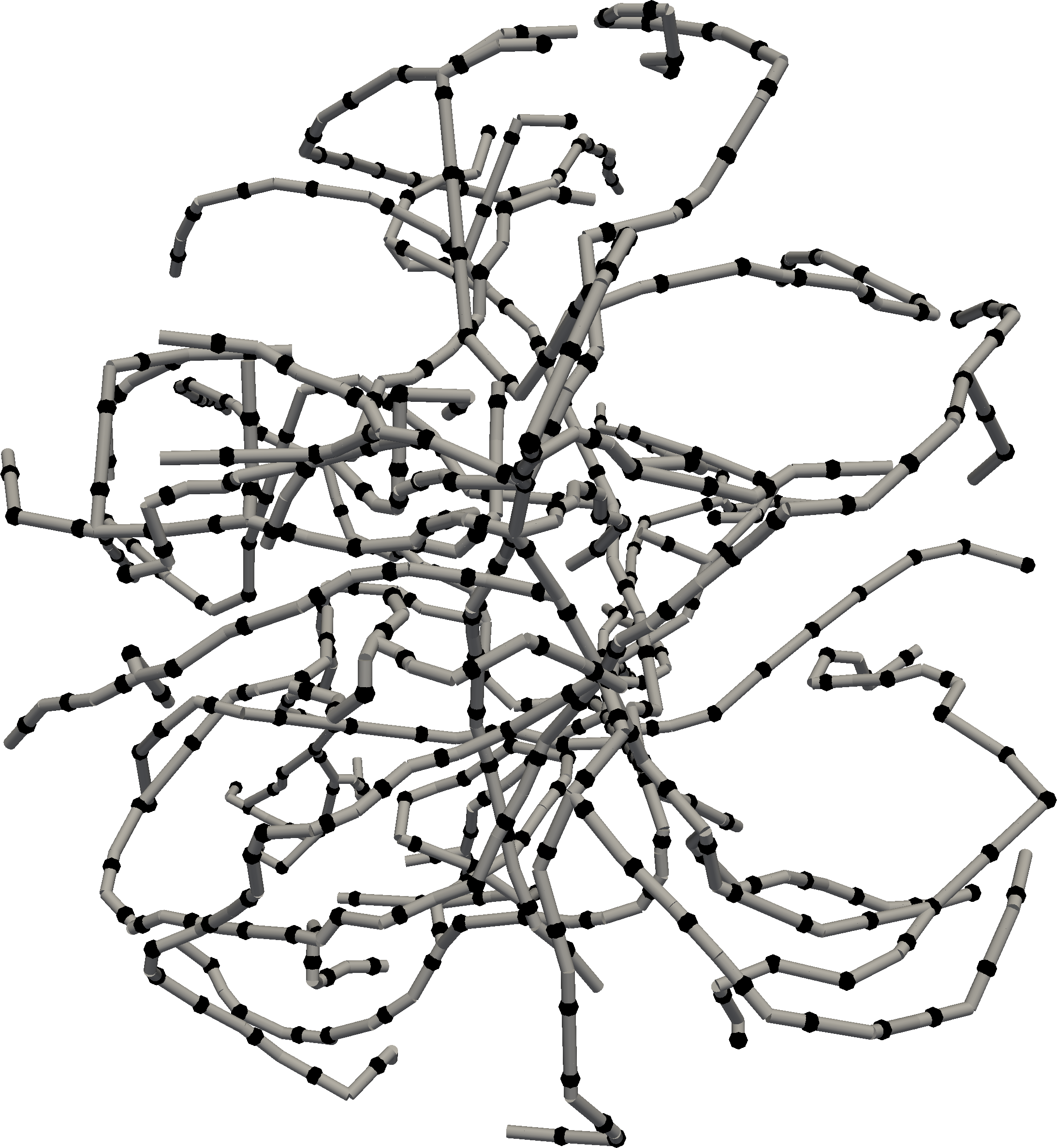}
  \hspace{0.1\textwidth}
  \includegraphics[height=0.24\textheight]{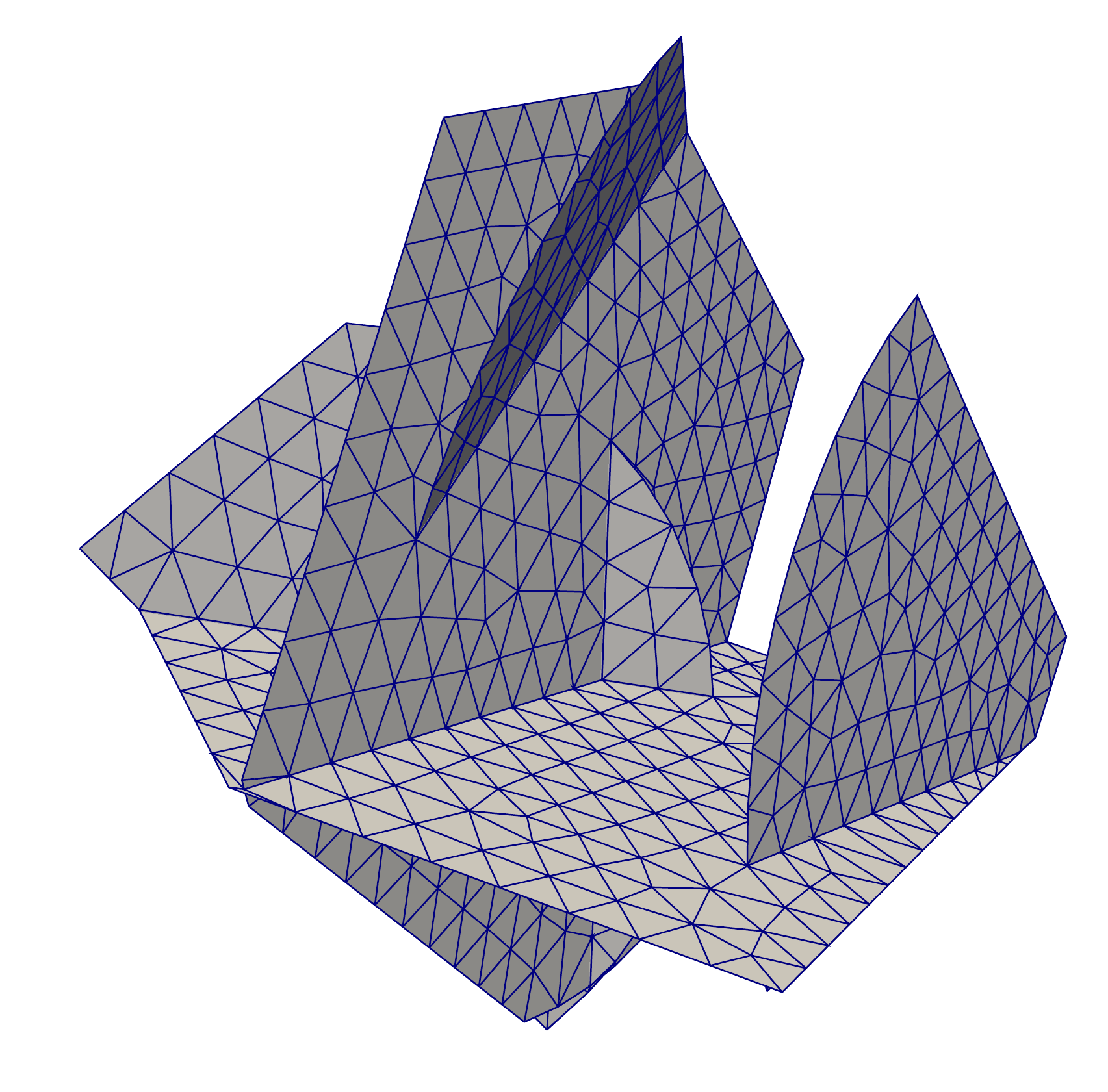}
 \end{center}

 \caption{One- and two-dimensional network grids}
 \label{fig:network_grids}
\end{figure}

As an additional difficulty, some boundary value problems are posed on domains $\Omega$ that
do not even have the structure of a topological manifold.  That is, not every point of $\Omega$
has a neighborhood that is homeomorphic to an open subset of $\mathbb{R}^d$.  Figure~\ref{fig:non_manifold_sets}
illustrates this: While the surface patch on the left is locally homeomorphic to Euclidean space,
the one in the middle is a T-junction, and the one on the right is a touching point.  Two-dimensional domains
with such features appear in applications like the simulation of closed-cell foams~\citep{nammi_myler_edwards:2010},
or networks of fractures in rock mechanics~\citep{mcclure_horne:2013,mcclure_babazadeh_shiozawa_huang:2015}.
Of considerable importance are also one-dimensional networks embedded into a two- or three-dimensional Euclidean space.
They appear in models of traffic networks~\citep{garavello_piccoli:2006},
supply chains~\citep{dapice_goettlich_herty_piccoli:2014},
but also in simulations of biological systems like root networks~\citep{Dunbabin2013}, neural networks~\citep{lang2011simulation},
or blood vessel networks~\citep{cattaneo2014computational}.
An overview over flow problems on networks is given in~\citep{bressan_canic_garavello_herty_piccoli:2014}.
Figure~\ref{fig:network_grids} shows two example domains for network problems.

Various discretization methods have been proposed for surface and network equations~\citep{dziuk_elliott:2013,olshanskii_reusken_grande:2009}.
Explicit discretizations, which are the focus of this work, use a grid of the same dimension as the domain.
For manifold surface grids it is reasonably simple to generalize grid data
structures to such a setting.  The main hurdles are admitting that the number of coordinates of a vertex
can be different from the effective grid dimension, and making sure that grids are not required to have a boundary.
Several standard simulation codes support such surface grids.  We mention Alberta~\citep{alberta}
(standalone and as part of \dune), AMDiS~\citep{vey_voigt:2007}, and FEniCS~\citep{rognes_ham_cotter_mcrae:2013}.
Moreover, the \geometrygrid \dune meta grid allows to embed any \dune grid into a Euclidean space of higher dimension.

Grid data structures for non-manifold grids are more challenging.  To handle T-junctions, for example, the data structure
must cope with the fact that element facets (i.e., edges in a 2d grid) may have more than two neighbors
(Figure~\ref{fig:non_manifold_sets}).  While this is not very difficult to implement,
it requires the introduction of additional data fields and logic, which is not used when the grid happens
to be a manifold.  Since the latter case is predominant, such additional features mean space and
run-time overhead for most users.  Therefore, standard grid data structures do not allow for non-manifold topologies.
Numerical examples of processes on non-manifold topologies in the literature are typically done using ad hoc implementations of the
necessary grid data structures.

Unfortunately, using ad hoc implementations wastes a lot of human resources.  While the domain may be non-standard, many of the
equations on network grids differ little from their Euclidean counterparts.  However, existing code like
finite element assemblers for diffusion and transport processes cannot be reused on ad hoc grid data structures.  They require detailed
knowledge of the grid, and it is a challenge to port an existing assembler to a new grid data structure.
Also, since the data structure is ad hoc, it is difficult to reuse it in other contexts.  In particular,
it is difficult to share it with other groups working in the same field.

The \dune software system%
\footnote{\url{www.dune-project.org}}~\citep{dunegridpaperI:08,dunegridpaperII:08}
has found an elegant solution for both problems.  \dune is a set of open-source C++ libraries dedicated to
various aspects of finite element and finite volume methods.
Its grid component, implemented in the \dunemodule{dune-grid} module, specifies
an abstract interface for computational grids.  The specification mandates what a data structure should be able
to do to qualify as a \dune grid, and how this functionality should be accessible from C++ code.
Examples of such functionality include being able to iterate over the elements and vertices, and getting
the maps from the reference element to the grid elements.
Those parts of a numerical simulation code that use the grid, such as the matrix assemblers or error estimators, are written
to use only the abstract grid interface.  Grid data structures implementing this interface are
then completely decoupled from the algorithms that use them.

This decoupling has various interesting consequences.
Using the \dune grid interface, it is easy to swap a given grid implementation for another one.  Indeed, in typical
\dune applications the C++ grid type is set once in the code, and handed around as a template parameter.
Changing the initial \cpp{typedef} and recompiling the code is usually sufficient to switch to an alternative grid data structure.
This possibility to easily switch between grid implementations allows to provide tailor-made grid data structures
for special simulation needs.  For example, \dunemodule{dune-grid} itself provides \yaspgrid, the implementation of a structured grid with
very little run-time and space overhead.
In contrast, \uggrid implements a very flexible unstructured grid with non-conforming and red--green refinement.

\bigskip

In this paper we present \foamgrid, a new implementation of the \dune grid interface that is dedicated to surface
and network grids.  Its main features are:
\begin{itemize}
 \item \foamgrid implements one- and two-dimensional simplex grids embedded in a physical Euclidean space
   of arbitrary dimension $w$.  Hence, it targets geometric and surface PDEs.
 \item The grids do not need to have the structure of a topological manifold.  Network configurations like the
   ones in Figures~\ref{fig:non_manifold_sets} and~\ref{fig:network_grids} are supported.
 \item A \foamgrid can be adaptively refined.  In the standard setup, refining a triangle results in four
   coplanar triangles.  Additionally, \foamgrid elements can be parameterized, i.e., they can be given a map
   that describes a nonlinear embedding of the element into $\mathbb{R}^w$.  As the element gets
   refined more and more, its shape approaches the one described by the parameterization
   (Figure~\ref{fig:refinement_with_parametrizations}).
 \item Finally, the domain of a \foamgrid can grow and shrink at run-time.  Elements can be added and removed
   even when there is no coarser ``father'' element, without invalidating the grid data structure.  Data can be
   transferred during this process.  This allows to elegantly simulate network growth and remodeling processes.
\end{itemize}

\foamgrid is available in a \dune module \dunemodule{dune-foamgrid},%
\footnote{\url{http://users.dune-project.org/projects/dune-foamgrid}}
and is installed just like any other \dune module.  \dunemodule{dune-foamgrid} is free software,
available under the either the LGPLv3+, or the GPLv2 with a linking exception clause.

Using \dune and \foamgrid for simulations of network and surface PDE problems has a number of important
advantages.  First of all, you do not have to implement the grid data structure yourself.  While not overly difficult, quite a bit
of thought has gone into \foamgrid, which would need work to be replicated.  Then,
since \foamgrid implements the \dune grid interface, large amounts of existing application
and infrastructure code can be used directly with \foamgrid.  This includes things like finite element
spaces and assemblers, error estimators, and grid file readers and writers.  This advantage is demonstrated in particular
by the numerical example
in Section~\ref{sec:example_fracture_network_flow}, which required no additional coding at all to extend an existing
planar code to a network setting.

As a further advantage, once  a user starts employing \foamgrid and the \dune grid interface, he immediately has
the power of all other \dune grid implementations at his disposal.
All of these are easily usable together with \foamgrid.  Hence, e.g., in simulations
that couple network grids with background grids,
several implementations of such background grids are readily available.%
\footnote{And the \dune extension module \dunemodule{dune-grid-glue} offers a convenient
way to do the coupling (\url{www.dune-project.org/modules/dune-grid-glue}).}
Since the \dune grid interface is a well-established standard, it is easy to learn how to
use \foamgrid. If a user already knows \dune, there is little additional knowledge
needed.
Since \foamgrid is open source, sharing code based upon it is particularly easy.

While \foamgrid has many interesting features, there is also a number of things it does not currently support.
For example, elements can only be simplices, and they must be one- or two-dimensional.  (The authors
could not think of a use case for a higher-dimensional network grid.  Write us if you know one.)
Adaptive grid refinement is currently non-conforming, which leads to hanging nodes, and, possibly,
to holes in the surface (Figure~\ref{fig:adaptive_parametrized_refinement}).  Finally, the current \foamgrid implementation
is purely sequential, and \foamgrid objects cannot be distributed across several processes.
However, the development of \foamgrid is ongoing, and these features may appear in later releases.

The present article is structured as follows.  Chapter~\ref{sec:grid_interface} briefly explains how to use \foamgrid.
Everything mentioned there is codified in the \dune grid interface specification,
and hence you can also read this chapter as an introduction to the use of the \dune grid interface.
Chapter~\ref{sec:additional_features} then explains the special features that \foamgrid offers.
In particular, these are support for T-junctions, adaptive refinement with element parameterizations, and the ability to ``grow''.
Finally, in Chapter~\ref{sec:numerical_examples} we give three numerical examples
showcasing the different features of \foamgrid.
The first shows unsaturated flow through a two-dimensional fracture network using finite elements. The second example shows $h$-adaptive, locally mass-conservative transport of a therapeutic agent in a microvascular network using finite volumes. The third one models
the growth of plant root networks.

\section{\foamgrid and the \dune grid interface}
\label{sec:grid_interface}

In this chapter, we start by describing the programmer interface of \foamgrid.  \foamgrid implements the \dune grid interface, hence in many central aspects, it can be used just like any other \dune grid.  This chapter focuses on these aspects. You can therefore also read it as a brief review of the \dune grid interface.  For more details, you may want to consult the \dune online documentation and \citep{dunegridpaperI:08,dunegridpaperII:08}.

The central class of the \foamgrid grid implementation is
\begin{c++}
template <int dim, int dimworld>
class FoamGrid;
\end{c++}
available from the header \file{dune/foamgrid/foamgrid.hh}.  This class implements a hierarchical grid as defined in
\citep[Def.\,13]{dunegridpaperI:08}, i.e., a coarse (or {\em macro}) grid, and element refinement trees rooted
in each of its elements.
The first template parameter \cpp{dim} is the grid dimension $d$, which must be either~1 or~2. The second template parameter \cpp{dimworld}
is the dimension $w$ of the Euclidean embedding space.  It must be equal to or greater than the grid dimension,
but can otherwise be arbitrary.  For the rest of this article we use \cpp{dim} and \cpp{dimworld} in code examples,
and $d$ and $w$ in text to denote the dimensions of the grid and the physical space, respectively.

To construct \foamgrid objects, the \cpp{FoamGrid} class implements the entire \dune grid interface for the
setup of unstructured grids.  In particular, the class \cpp{GridFactory} is implemented for \cpp{FoamGrid}.
Thus, all file-reading methods based on this interface are available.
For example, files in the \gmsh format~\citep{geuzaine_remacle:2015} can be read by using the line
\begin{c++}
std::shared_ptr<FoamGrid<2,3>  grid( GmshReader<FoamGrid<2,3>>::read("filename.msh") );
\end{c++}
This will read the file named ``\file{filename.msh}'', and set up a new \cpp{FoamGrid<2,3>} object with it.
The new grid object is returned in a shared pointer called \cpp{grid}.  Note that vertex coordinates
in \gmsh files always have three components, so reading a \gmsh file into a \cpp{FoamGrid<1,2>} object
will discard the third entry.
As a special feature, \gmsh files can contain elements with polynomial geometries of order up to five.
While \foamgrid element geometries are always affine, \foamgrid can use the higher-order geometries during
mesh refinement (Section~\ref{sec:element_parametrizations}).

Writing \cpp{FoamGrid} objects to disk is equally straightforward. All \dune file writing codes rely on the grid
interface only, and can therefore be used with \foamgrid. For example, writing the object pointed to by the \cpp{grid} shared pointer
into a VTK file called \file{my\_filename.vtu} can be achieved by including the header \file{dune/grid/io/file/vtk.hh}
from the \dunemodule{dune-grid} module, and writing
\begin{c++}
typedef FoamGrid<2,3> GridType;
VTKWriter<GridType::LeafGridView> vtkWriter(grid->leafGridView());
vtkWriter.write("my_filename");
\end{c++}
The resulting file can be visualized, e.g., with the \paraview software.

\subsection{Elements and geometries}
\label{sec:elements_and_geometries}

In \dune, finite element assembly typically takes place on the leaf elements of the refinement trees.  These elements
form the {\em leaf grid view}, which encapsulates the notion of a textbook non-hierarchical finite element grid.
From a \foamgrid, the leaf grid view can be obtained in the usual way, i.e.,
\begin{c++}
auto foamGridLeafView = grid->leafGridView();
\end{c++}
which already has been used in the VTK example above.

Access to grid elements and vertices is provided by the grid view.
Elements can be iterated over using \cpp{begin}/\cpp{end}-iterators
\begin{c++}
for (auto it = foamGridLeafView.begin<0>();
     it != foamGridLeafView.end<0>();
     ++it)
{
  // do something with the element in '*it'
}
\end{c++}
where the number \cpp{0} specifies that the loop is to be over the grid elements (it is the codimension of the grid elements
with respect to the grid).  Using the C++11 range-based-\cpp{for} syntax, the same loop can be written more concisely
\begin{c++}
for (const auto& element : elements(foamGridLeafView))
{
  // do something with the element in 'element'
}
\end{c++}
Similarly, a loop over all vertices of the grid is written as
\begin{c++}
for (auto it = foamGridLeafView.begin<dim>();
     it != foamGridLeafView.end<dim>();
     ++it)
{
  // do something with the vertex in '*it'
}
\end{c++}
In this code, the number \cpp{dim} (i.e., the grid dimension), specifies that the loop is to be over the grid vertices,
because vertices are zero-dimensional and hence have codimension \cpp{dim} in a \cpp{dim}-dimensional grid.
Alternatively, one can write
\begin{c++}
for (const auto& vertex : vertices(foamGridLeafView))
{
  // do something with the vertex in 'vertex'
}
\end{c++}

The objects \cpp{vertex} and \cpp{element} (or \cpp{*it} in the iterator loops) are instances of what in \dune terminology
are called {\em entities}.  Entities are implemented by the \cpp{Entity} interface class in \dunemodule{dune-grid}.
They provide topological information about the grid elements and vertices, like links to the corners
vertices of an element, and to the father and descendant elements in the refinement tree.  In all these aspects, \cpp{FoamGrid}
objects behave just like any other \dune grids.  Furthermore, each element entity provides a \cpp{Geometry} object,
which represents the affine map $F : T_\text{ref} \to T \subset \mathbb{R}^w$ from the reference element $T_\text{ref}$ to the grid element $T$.
This map provides the geometrical information needed to assemble finite element and finite volume systems, like evaluation
of $F$ and its inverse $F^{-1}$, the inverse transposed Jacobian matrix $\nabla F^{-\text{T}}$, and the functional
determinant $J \coloneqq \sqrt{\det \nabla F^\text{T} \nabla F}$.
Note that since for surface grids the world dimension $w$ is larger than the dimension of the reference element $T_\text{ref}$,
the image of $T_\text{ref}$ under $F$ is a set of measure zero in $\mathbb{R}^w$.  In finite-precision arithmetic the argument of
the method implementing $F^{-1}$ will typically not be in the domain of $F^{-1}$.
The \cpp{Geometry} implementation of \foamgrid therefore extends $F^{-1}$ to the entire space $\mathbb{R}^w$.
Given a point $x \in \mathbb{R}^w$
not necessarily in $T$, \cpp{FoamGrid} computes a point $\xi$ in the plane spanned by $T_\text{ref}$
such that $|F(\xi)-x|$ is minimized.  The affine function $F$ is tacitly extended from $T_\text{ref}$
to its entire affine hull here.

To compute element fluxes, elements in a \dune grid provide a set of so-called {\em intersections}, which relate
elements to their neighbors.  This mechanism is very flexible, and in particular handles general non-conforming
situations very well.  Nevertheless, using grids for network domains stretches the bounds of the current intersection concept,
and some generalization is needed.  We have therefore dedicated a separate chapter to \foamgrid intersections
(Chapter~\ref{sec:intersections}).

\subsection{Attaching data to grids}
\label{sec:attaching_data}

Data is attached to \foamgrid objects in the same way as to other \dune grids.  Each grid view object can
provide a corresponding \cpp{IndexSet} object
\begin{c++}
const auto& indexSet = foamGridLeafView.indexSet();
\end{c++}
This index set provides an integer number for each entity (i.e., vertex, edge, or element) of the grid view.  For each dimension and reference element
type, these numbers are consecutive and start at zero.  They can hence be used to address random-access
containers holding the simulation data.  This approach is convenient, flexible, and efficient.

Data stored in arrays is lost if the grid changes, either by refinement (Section~\ref{sec:adaptive_refinement})
or by grid growth (Section~\ref{sec:grid_growth}).  To preserve data across grid modifications, \foamgrid, just like
any other \dune grid, additionally provides a set of persistent numbers.  These are obtained by an \cpp{IdSet} object
\begin{c++}
const auto& idSet = grid->localIdSet();
\end{c++}
(remember that in our initial example the variable \cpp{grid} was a shared pointer to a \cpp{FoamGrid}).  Persistent numbers are neither consecutive
nor restricted to start at zero, but they can be used to access search trees or hash maps.  Before modifying the grid,
all simulation data must be copied into such data structures, and copied back to arrays after the modification
is completed.  While such copying is costly, its run-time is usually negligible compared to the cost of the
actual grid modification.

\subsection{Adaptive refinement}
\label{sec:adaptive_refinement}

\foamgrid supports red refinement (non-conforming refinement) of simplices, where each triangle is split into four congruent
smaller triangles.  If a two-dimensional \foamgrid is refined locally, then hanging nodes appear in the grid.  Depending
on the discretization used, this may or may not be a problem.  True red--green refinement, which avoids the hanging nodes,
may appear in later versions of \foamgrid.

Adaptive grid refinement in \foamgrid is
controlled via the standard \dune grid interface. In a first step the method
\begin{c++}
bool mark (int refCount, const Codim<0>::Entity& element);
\end{c++}
is used to mark an element \cpp{element} for refinement (\cpp{refCount} $> 0$) or coarsening (\cpp{refCount} $< 0$).
The method returns \cpp{true} if the element was successfully marked. The mark of an element \cpp{element} can be obtained with the method
\begin{c++}
int getMark (const Codim<0>::Entity& element) const;
\end{c++}
which returns \cpp{1} for elements marked for refinement, \cpp{-1} for elements marked for coarsening, and \cpp{0}
for unmarked elements.

The grid is then modified in a second step with the methods
\begin{c++}
bool coarsen = grid.preAdapt(); // true if at least one element
                                // will be coarsened
bool refined = grid.adapt();    // true if at least one element
                                // was refined
grid.postAdapt();
\end{c++}
Between \cpp{preAdapt()} and \cpp{adapt()} it is possible to check the following flag
\begin{c++}
bool mightVanish = element.mightVanish(); // true if the element might
                                          // vanish due to coarsening
                                          // by grid.adapt()
\end{c++}
Similarly, between \cpp{adapt()} and \cpp{postAdapt()} one can check the flag
\begin{c++}
bool isNew = element.isNew(); // true if element was created by last
                              // call to adapt()
\end{c++}
Both are useful to manage the transfer of data associated with the grid. As \foamgrid itself does not store any
associated data, the data transfer from the old grid to the adapted grid is managed by the user. An example
featuring adaptive refinement and coarsening is presented in Section~\ref{sec:bloodflow}.

\section{Additional features}
\label{sec:additional_features}

The previous chapter has described those aspects where \foamgrid behaves just like any other \dune grid.
However, \foamgrid also has a few features that
set it apart from most other \dune grids.  The present chapter is dedicated to those features.

\subsection{Handling intersections in a non-manifold grid}
\label{sec:intersections}

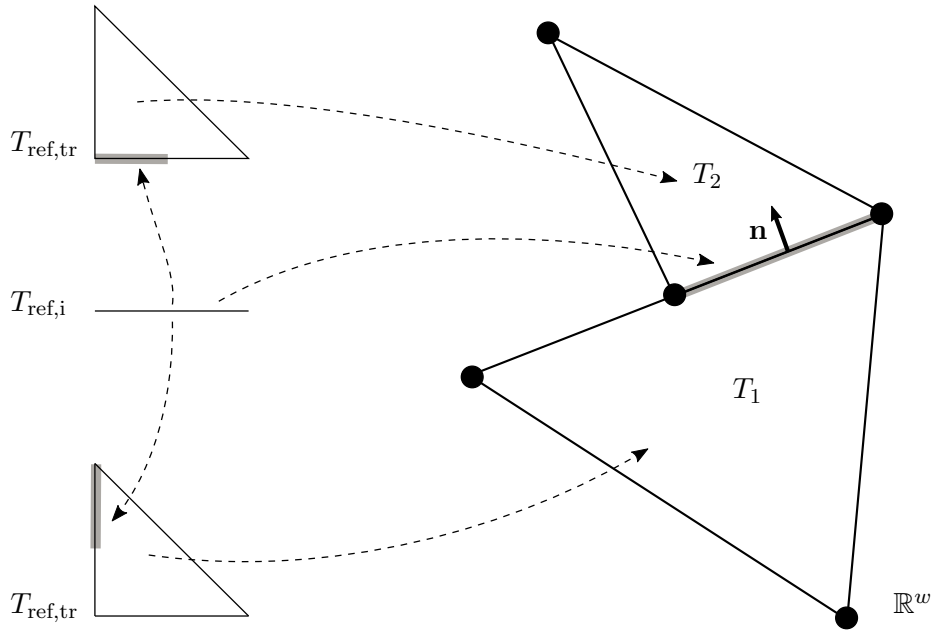
\begin{figure}
 \begin{center}
  \begin{overpic}[width=0.65\textwidth]{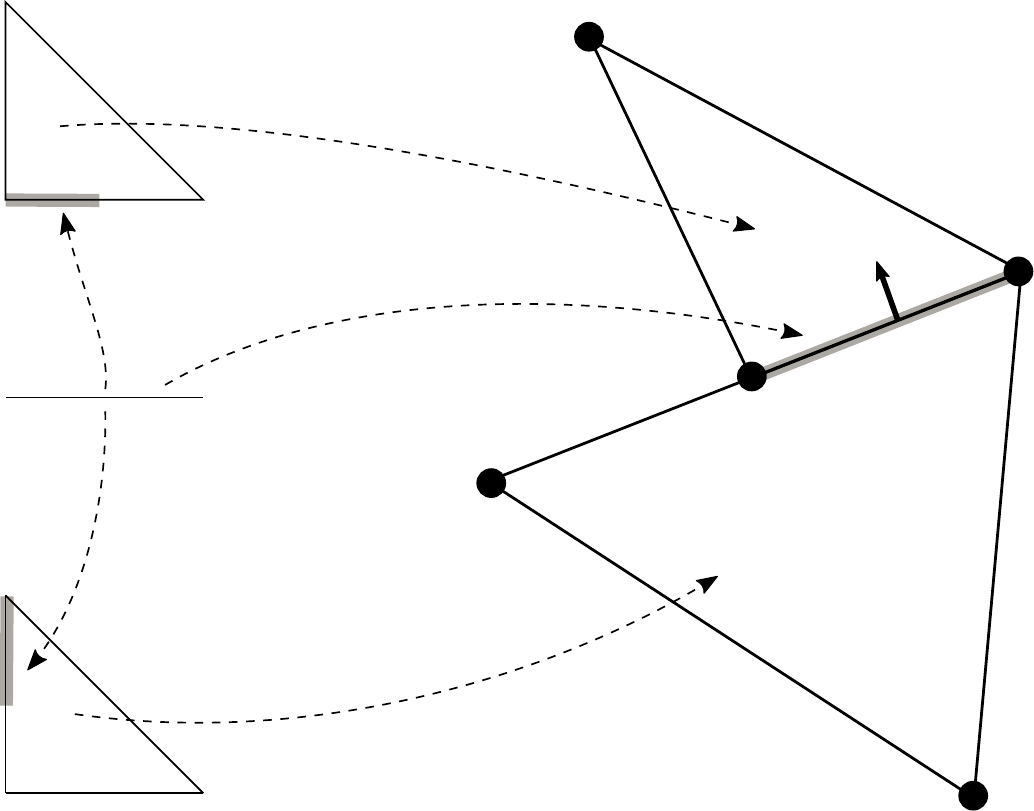}
   \put(80,29){$T_1$}
   \put(75,56){$T_2$}

   \put(82,49){$\mathbf{n}$}

   \put(-10,2) {$T_\text{ref,tr}$}
   \put(-10,40){$T_\text{ref,i}$}
   \put(-10,60){$T_\text{ref,tr}$}
   
   \put(100,2){$\mathbb{R}^w$}
  \end{overpic}
 \end{center}

 \caption{Intersection between two elements $T_1$ and $T_2$.  From its reference element $T_{\text{ref},\text{i}}$,
    there are two maps to $T_\text{ref,tr}$ (the reference element of $T_1$ and $T_2$), and one to $\mathbb{R}^w$, which describe the shape of the intersection in
    $T_1$, $T_2$, and $\mathbb{R}^w$, respectively.}
 \label{fig:dune_intersection}
\end{figure}

The defining feature of \foamgrid is its capability to handle network grids, i.e., grids with a
non-manifold topology (Figures~\ref{fig:non_manifold_sets} and~\ref{fig:network_grids}).  In particular, more than two
elements can meet in a common vertex in a one-dimensional grid, and more than two triangles can meet in a common edge
in a two-dimensional grid.

Information about how a given element relates to its neighbors is essential
for finite volume and DG methods, or more generally all methods involving element boundary fluxes.
For this, the \dune grid interface provides the notion of
{\em intersections}.  In \dune terminology, an {\em intersection} is the set-theoretic intersection between the closures
of two neighboring elements.  Only intersections that have positive $(d-1)$--dimensional measure qualify as intersections
in the \dune sense,
and those are equipped with a coordinate system, i.e., a map from a $(d-1)$--dimensional reference element $T_{\text{ref},\text{i}}$ to the intersection
(Figure~\ref{fig:dune_intersection}).
Note that if the grid is conforming, then intersections will be geometrically the same as shared element faces.
However, in general non-conforming grids, intersections are only subsets of element faces.

Intersections provide all information needed to compute element boundary fluxes.  In particular, for each point
on an intersection, you can get the vector $\mathbf{n}$ that is tangent to the element at that point,
and normal to the element boundary.  Also, you get the embeddings of the intersection into the two elements,
in form of maps from the intersection reference element $T_{\text{ref},\text{i}}$ to the element reference elements $T_{\text{ref},\text{tr}}$ (Figure~\ref{fig:dune_intersection}).
Those maps pick up the same ideas used to model element geometries in Section~\ref{sec:elements_and_geometries}.
They are called {\em geometry-in-inside} and {\em geometry-in-outside}, respectively.

In C++ code, {\em intersections} are objects of type \cpp{Intersection}.
For any given element (which is then called the {\em inside} element), all its intersections with neighboring elements can be accessed
by traversing them with a dedicated iterator.  Using range-based \cpp{for} syntax, a loop over all intersections
of the element called \cpp{element} is written as
\begin{c++}
for (const auto& intersection : intersections(foamGridLeafView,element))
{
  // do something with 'intersection'
}
\end{c++}
See the \dunemodule{dune-grid} class documentation for the details on the user interface of the \cpp{Intersection}
class.

Unfortunately, the \dune intersection mechanism, as flexible as it is, is not flexible enough for network grids.
The original idea was that while elements may have more than one neighbor across a given facet, there is
(even in non-conforming situations) at most one neighbor {\em at any given point} on that facet.
This reflects the assumption that computational domains are expected to be topological manifolds.
At the time of writing, this assumption is still reflected in the grid interface.  In particular, the method
\begin{c++}
bool neighbor() const;
\end{c++}
(a public member of the \cpp{Intersection} class),
informs whether there is a neighbor across a given intersection.  However, this information is insufficient in network grids,
where even in a conforming grid there may be an arbitrary number of neighbors meeting at a common intersection.
Finite volume and DG methods need access to this group of neighbors, to know how to distribute the flux
across this intersection.

For this reason the intersection concept and programmer interface is being revisited, and is likely to undergo changes in the future
to better support network grids.  Changes to the \dune interface can only be made in a democratic process, and it is
therefore unclear at what point in time such a change will happen.

Two different approaches have been proposed to the \dune grid development community.  We describe them both briefly here.
The full proposal text is available at \url{www.dune-project.org/modules/dune-foamgrid}.

The first approach tries to be as minimally invasive as possible.  In particular, it retains the notion of an intersection
as an object that relates {\em two} elements.
The extension consists of two semantic rules, and a change to the \cpp{neighbor} method.  The first of the semantic
rules makes sure that the ``number of neighbors'' across a given intersection is well-defined.
Remember that the {\em geometry-in-inside} is, roughly speaking, the intersection interpreted as
a subset of the element $T_1$.
\begin{interfacerule}
For any two intersections of a given element $T_1$, the geometries-in-inside are either disjoint or identical.
\end{interfacerule}
With the number of neighbors properly defined, the \cpp{neighbor} method is generalized to return this number
instead of a yes/no answer:
\begin{c++}
std::size_t neighbor() const;
\end{c++}
Note that this change is fully backward-compatible, as intersections in a non-network grid will return either \cpp{1} or \cpp{0}
here, which casts to the values \cpp{true} or \cpp{false} as used previously.

The second semantic rule provides a way to find all intersections that share a common geometry-in-inside.
No additional interface method is added for this.
Rather, it is guaranteed that all such intersections appear consecutively when traversing the intersections with the
intersection iterator.
\begin{interfacerule}
 If more than one neighbor is reachable over a given geometry-in-inside, then all intersections for this
 geometry-in-inside shall be traversed consecutively by the intersection iterator.
\end{interfacerule}
With this rule, groups of neighbors can be identified and used in flux computations.

\bigskip

The second approach is more radical, because it changes the idea of an intersection itself.
Intersections cease to be objects that relate {\em pairs} of elements.  Rather, they now become
objects that relate {\em groups} of elements.
As a consequence, each intersection still has only one geometry-in-inside. However, for each intersection
there is now more than one outside element, each with corresponding geometry-in-outside and index-in-outside.

To access this information, more interface methods need to be changed.
First of all, the \cpp{neighbor} method needs to be changed as proposed above.  Secondly the methods
\begin{c++}
Entity outside () const;
LocalGeometry geometryInOutside () const;
int indexInOutside () const;
\end{c++}
of the \cpp{Intersection} interface class need to be replaced by
\begin{c++}
Entity outside (std::size_t i=0) const;
LocalGeometry geometryInOutside (std::size_t i=0) const;
int indexInOutside (std::size_t i=0) const;
\end{c++}
respectively.  Rather than returning the unique outside element or its geometry or index, the methods
must now return the corresponding quantity for the \cpp{i}-th outside element.

These changes are again fully backward-compatible.  In grids without multiple intersections,
at most the $0$-th outside element will be available.  The default parameter ensures that this intersection
will be returned when the method is called without argument.

As an advantage, this proposal retains the rule that the geometries-in-inside must form a disjoint partition of the element boundary
(modulo zero-sets).  Also, it is easier to attach data to such intersections, which is a feature that has been
requested various times in the past.  On the downside, to iterate over all neighbors of an element,
two nested loops are needed, instead of only one.  This will make some code a bit longer, and more
difficult to read.

Both proposals are currently under discussion.  However, even with
the current status quo, applications involving fluxes can be written for network domains.  One-dimensional domains
are straightforward as there the intersections are a fortiori conforming (see Sections~\ref{sec:bloodflow} and~\ref{sec:root_networks}).
Two-dimensional networks need more trickery, but can also be made to work.  Once either of the proposed interface
extensions has been officially accepted, implementations of such methods will be much simpler.

\subsection{Element parametrizations}
\label{sec:element_parametrizations}

\begin{figure}
 \begin{center}
 \includegraphics[width=0.32\textwidth]{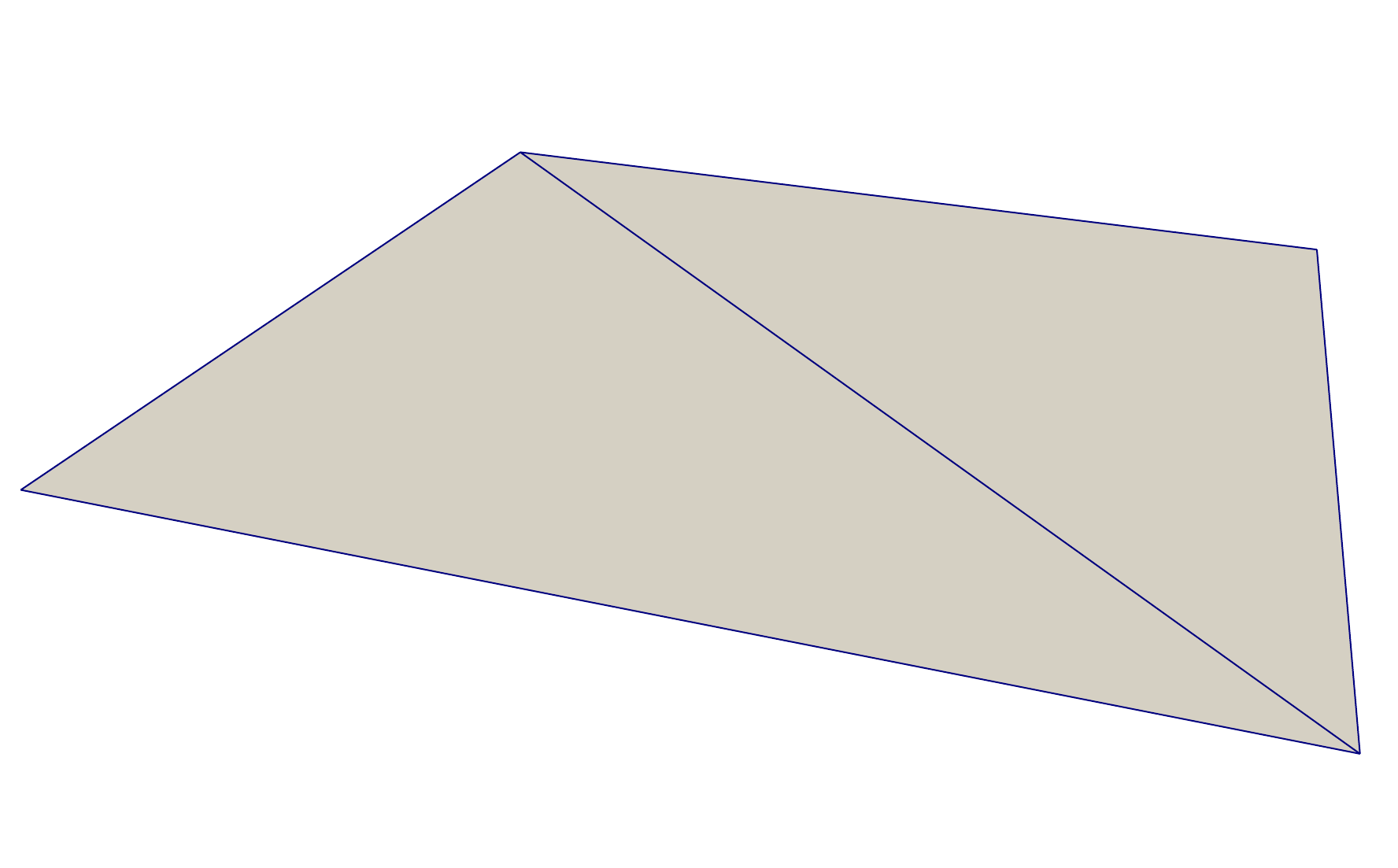}
 \includegraphics[width=0.32\textwidth]{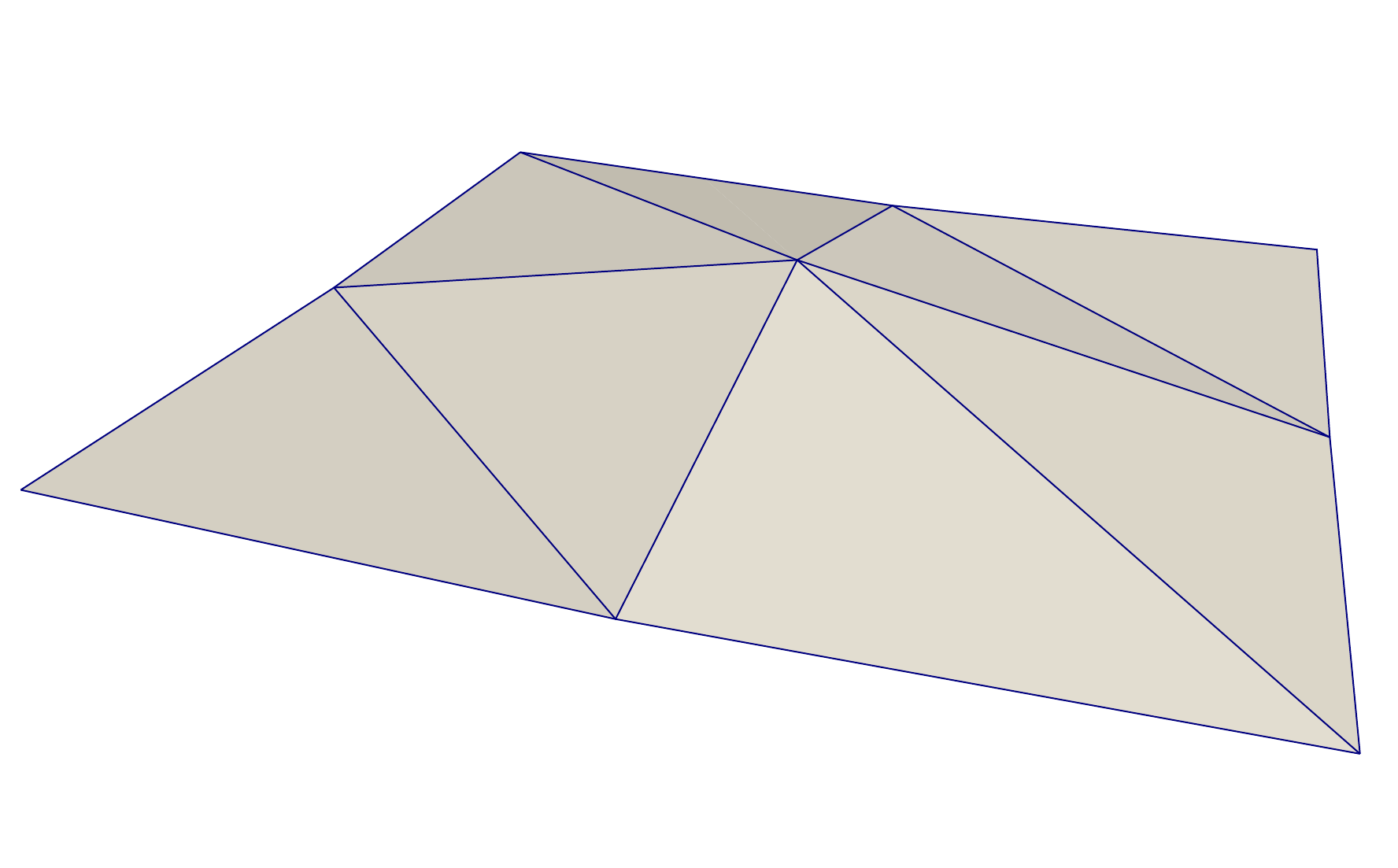}
 \includegraphics[width=0.32\textwidth]{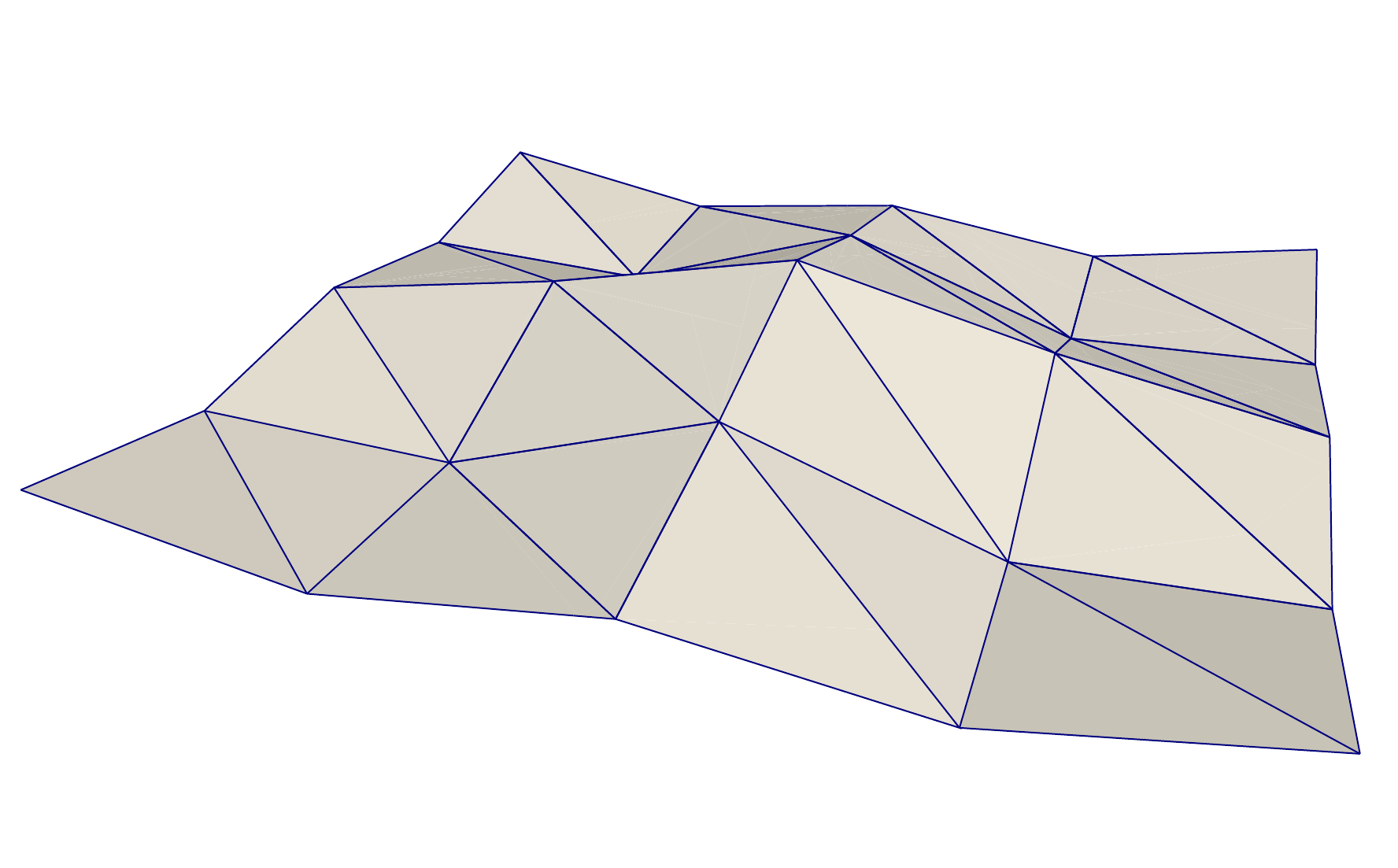}

 \includegraphics[width=0.32\textwidth]{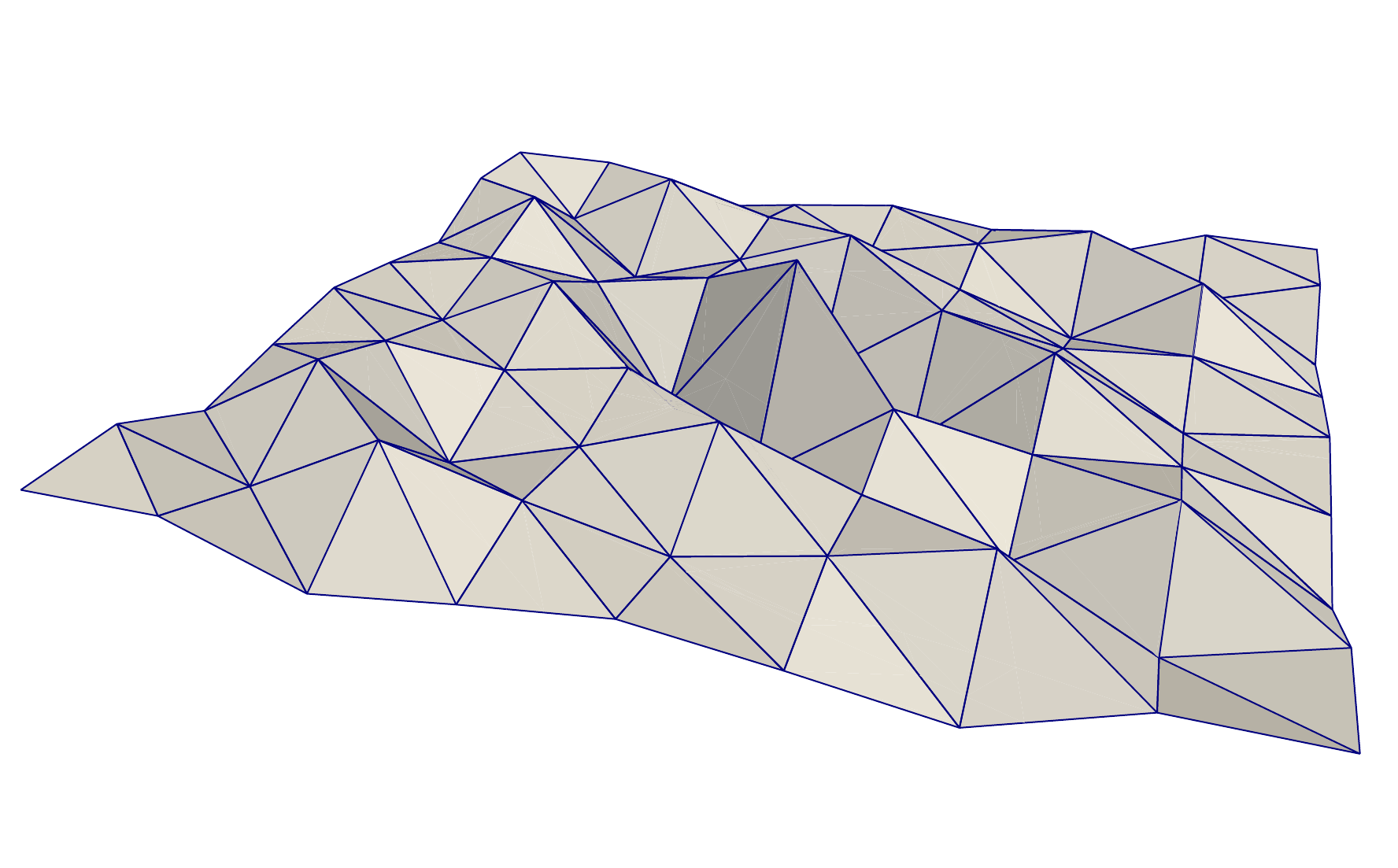}
 \includegraphics[width=0.32\textwidth]{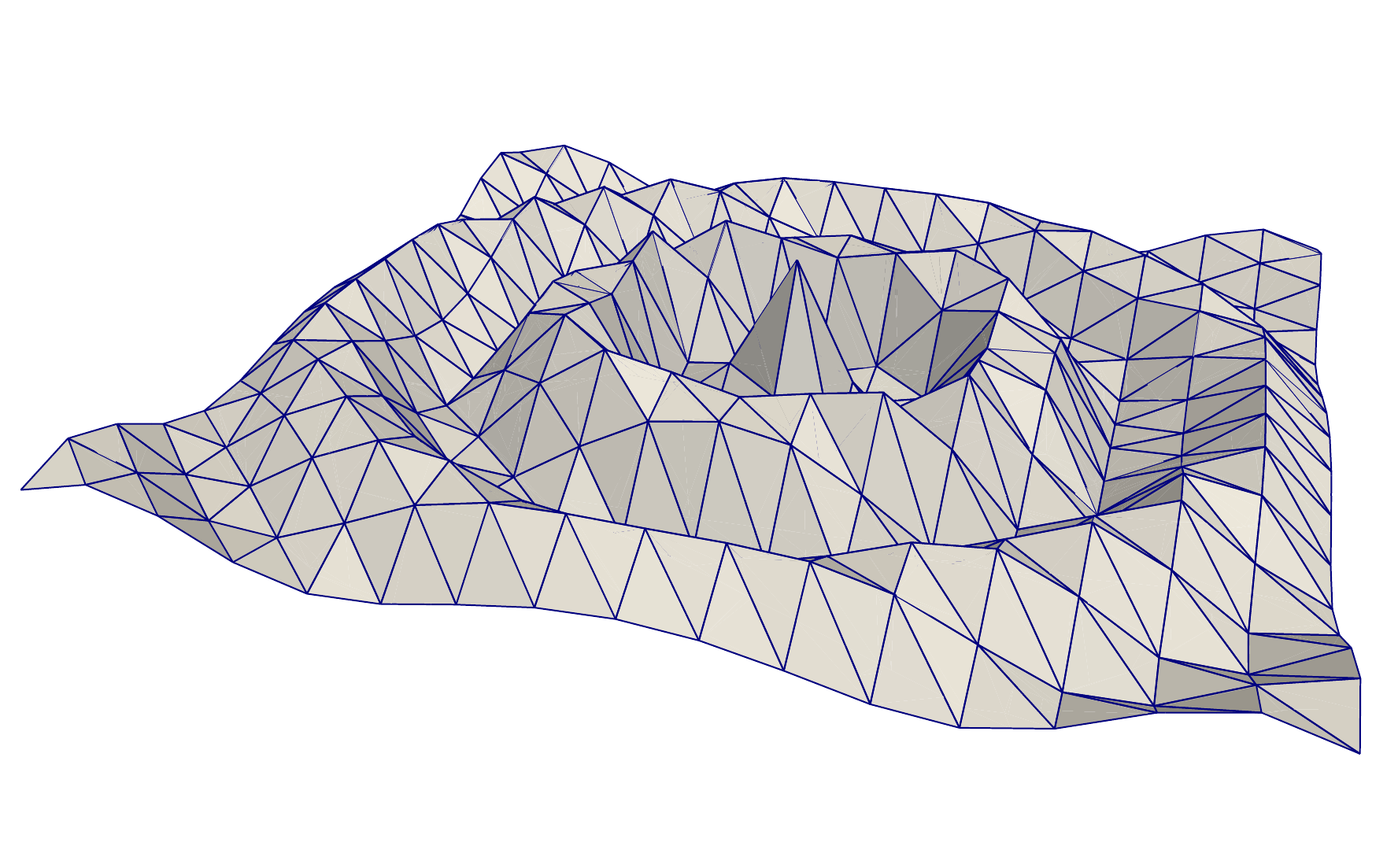}
 \includegraphics[width=0.32\textwidth]{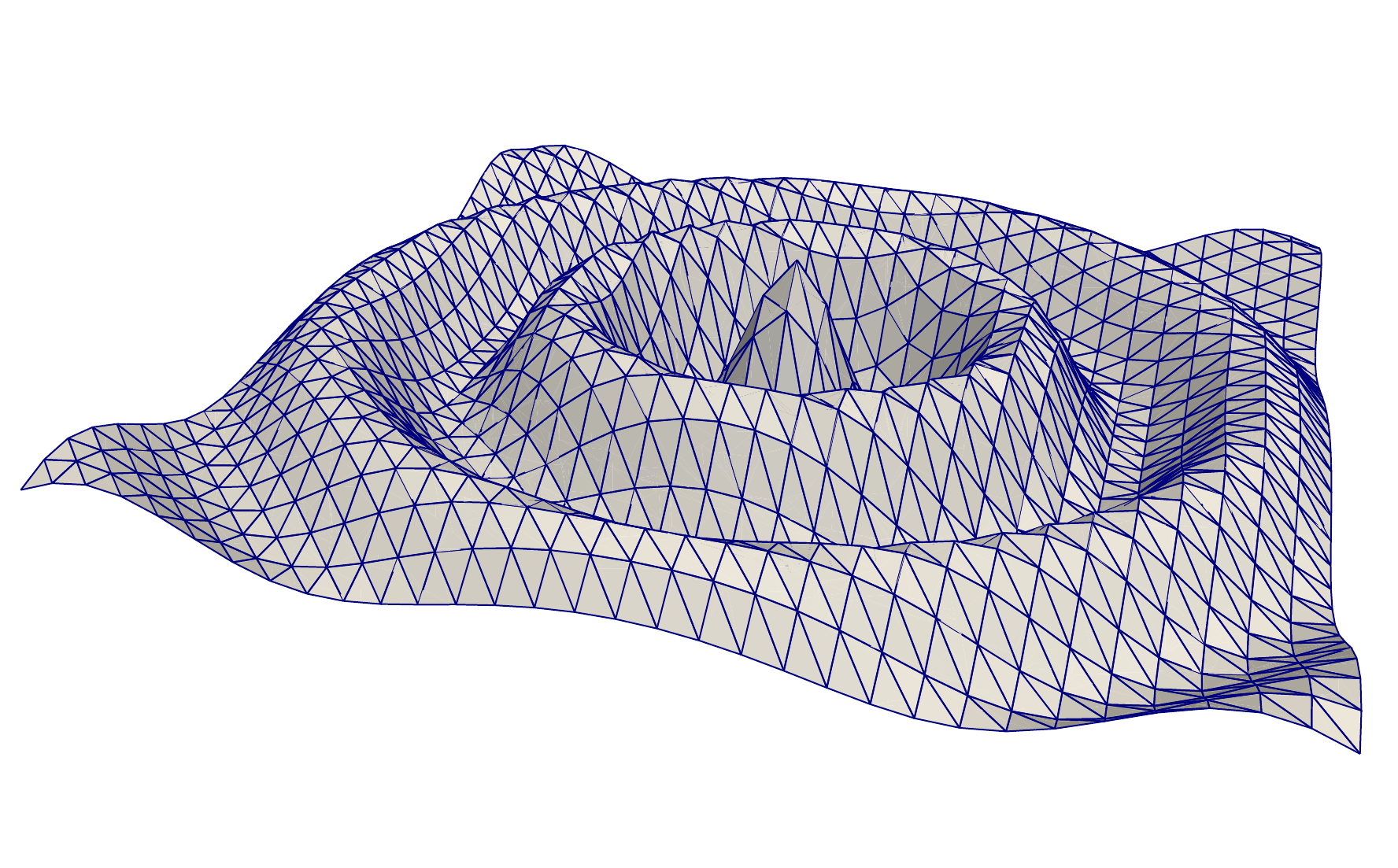}
 \end{center}
 \caption{Grid refinement with element parametrizations}
 \label{fig:refinement_with_parametrizations}
\end{figure}

In a standard non-conforming red refinement algorithm, new vertices are placed at the edge midpoints of refined
elements.  That way, while the grid gets finer and finer, the geometry of the grid remains identical to the
geometry of the coarsest grid.  To also allow improvements to the geometry approximation, \foamgrid can
use element parametrizations.
Each coarse grid element $T$ with reference element $T_\text{ref}$ of a \foamgrid can be given a map
\begin{equation*}
 \varphi_T : T_\text{ref} \to \mathbb{R}^w,
\end{equation*}
which describes an embedding of $T$ into
physical space $\mathbb{R}^w$.  This does not influence the grid itself --- elements of a \foamgrid are always affine.
However, when refining the grid, new elements are not inserted at edge midpoints.  Rather, for each new vertex
which would appear at an edge midpoint in standard refinement, the corresponding local
position in its coarsest ancestor element $T$ is determined using the refinement tree.  This position
is then used as an argument for $\varphi_T$, and the result is used as the position of the new vertex.
That way, the grid approaches the shape described by the parametrization functions $\varphi_T$ more and more as the
grid gets refined (Figure~\ref{fig:refinement_with_parametrizations}).

In \dunemodule{dune-grid},
element parametrizations are implemented as small C++ objects.  These must inherit from the abstract base class
\begin{c++}
VirtualFunction<FieldVector<double,dim>,
                FieldVector<double,dimworld> >;
\end{c++}
declared in \file{dune/common/function.hh}.%
\footnote{For various reasons there is discontent with using the \cpp{VirtualFunction} class to implement element
  parametrizations.  Therefore, it is expected to be replaced by something else eventually.  However, at the time
  of writing, there is no specific proposal for such a replacement.  In any case, while the details of the
  implementation are under discussion, the general idea of element parametrizations is not disputed, and can be
  expected to remain the way it is described here.
}
One object of this type needs to be created for each
element of the coarse grid.  The base class has a single pure virtual method
\begin{c++}
virtual void evaluate(const FieldVector<double,dim>& x,
                      FieldVector<double,dimworld>& y) const = 0;
\end{c++}
which implements the evaluation of $\varphi_T$.  The first argument \cpp{x} is a position in local coordinates
of the corresponding coarse grid triangle.  The result $\varphi_T(x)$ is returned in the second argument \cpp{y}.

Element parametrizations are handed to the \cpp{GridFactory} during grid construction.
Normally, grid elements are entered using the method
\begin{c++}
void insertElement(const GeometryType& type,
                   const std::vector<unsigned int>& vertices);
\end{c++}
of the \cpp{GridFactory} class.
For parametrized elements, there is the alternative method
\begin{c++}
void insertElement(const GeometryType& type,
                   const std::vector<unsigned int>& vertices,
                   const std::shared_ptr<VirtualFunction<
                                           FieldVector<ctype,dim>,
                                           FieldVector<ctype,dimworld> 
                                        > >& elementParametrization);
\end{c++}
This inserts the element with vertex numbers given in the array \cpp{vertices} and an element parametrization
given by the object \cpp{elementParametrization} into the grid.
The \cpp{GmshReader} uses this method for some of the higher-order elements that can appear in \gmsh grid files.

To see the effect of element parametrizations,
Figure~\ref{fig:refinement_with_parametrizations} shows an example where the coarsest grid consists of only
two triangles covering the domain $\Omega = [-1,1]^2 \times \{0\}$.
As a parametrization we use the global function
\begin{equation}
\label{eq:example_parametrization}
\varphi : \mathbb{R}^2 \to \mathbb{R}^3,
\qquad
\varphi \begin{pmatrix}x_1\\ x_2\end{pmatrix}
=
\begin{pmatrix}x_1\\x_2\\0.2\cdot \exp (-\abs{x})\cos( 4.5\pi \abs{x})\end{pmatrix},
\end{equation}
and each element parametrization $\varphi_T$ first maps its local coordinates to $\mathbb{R}^2$,
and then applies the global function $\varphi$.
Hence, upon refinement, the grid will approach the
graph of the function~\eqref{eq:example_parametrization}.  The code for this example is provided in the \dunemodule{dune-foamgrid} module
itself, in the file \file{dune-foamgrid/examples/parametrized-refinement.cc}.

\begin{figure}
 \begin{center}
  \includegraphics[width=0.7\textwidth]{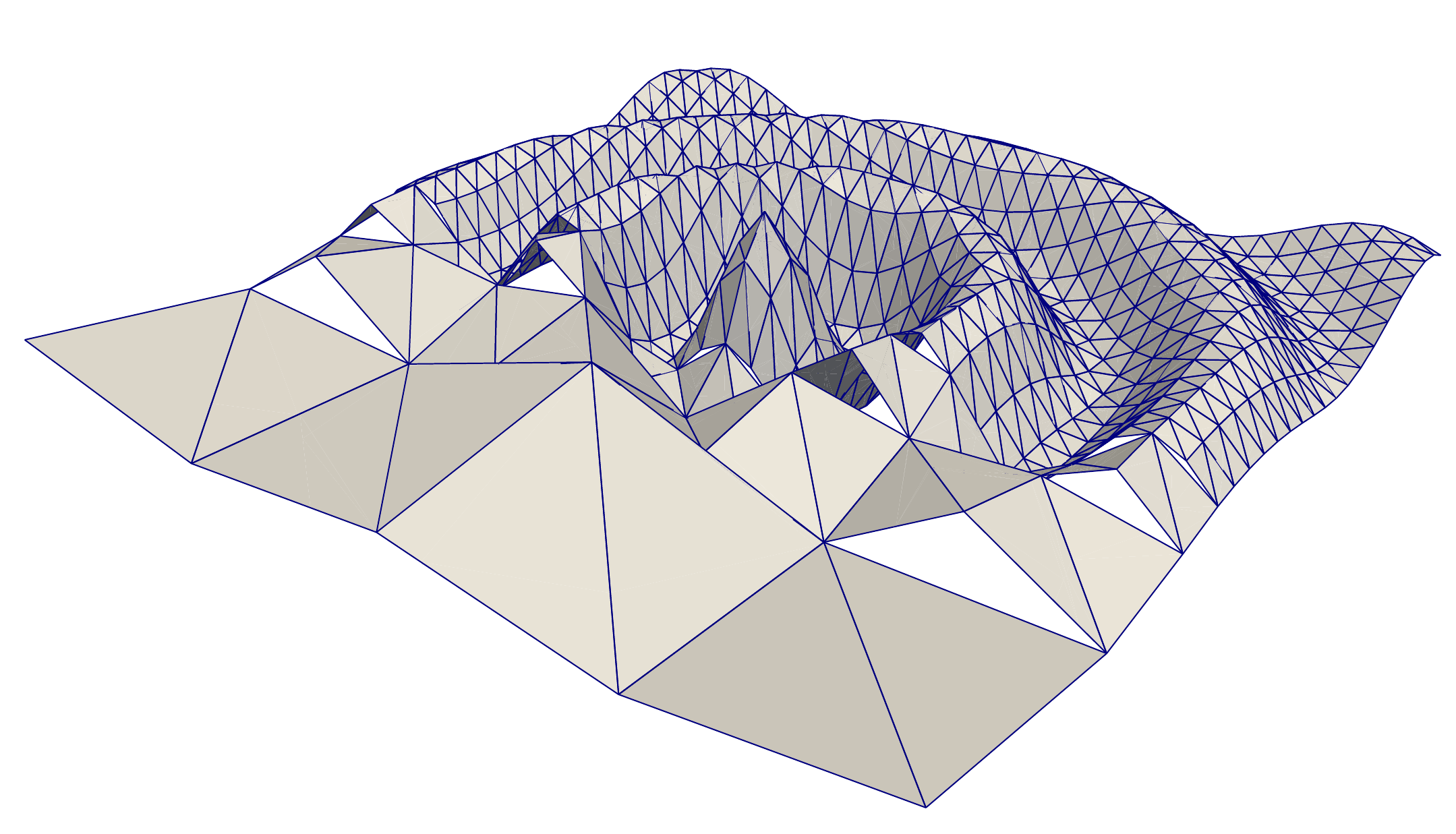}
 \end{center}
 \caption{Adaptive refinement with element parameterizations leads to non-conforming geometries}
 \label{fig:adaptive_parametrized_refinement}
\end{figure}

There is one pitfall when using element parametrizations together with non-conforming adaptive refinement
for two-dimensional grids. If a hanging node appears in the course of refinement, the position of this node
is determined by the parametrization functions of the corresponding element, which is typically {\em not}
the midpoint of the adjacent longer edge.  As a consequence, the grid will have holes wherever different refinement levels meet (Figure~\ref{fig:adaptive_parametrized_refinement}).  While this may seem surprising at first sight,
it is nevertheless a logical consequence of the hanging nodes refinement.  The continuous domain
is approximated by discontinuous grids, in direct correspondence to how discontinuous FE functions
may be used to approximate continuous PDE solutions.

The non-conforming geometry approximation shows another shortcoming of the current \dune intersection
concept.  There, intersections are defined as set-theoretic intersections of (closures of) neighboring
elements.  However, in the geometrically non-conforming situation, neighboring elements do not actually
intersect, and one can only speak of logical intersections.  The practical consequence of this is that
intersections do not have a single well-defined shape in $\mathbb{R}^w$ anymore.  Rather, there are now {\em two}
of them.  This possibility to have two global element shapes is not currently reflected by the \dune
grid interface.  In the current implementation of \foamgrid, the method \cpp{Intersection::Geometry}
will always return the global shape of the intersection as seen from the inside element.

\subsection{Grid growth}
\label{sec:grid_growth}

A certain number of network problems is posed on domains that grow and/or shrink in the course of the simulation.
Examples are fracture growth processes and simulations of bone trabeculae remodeling.  To support such simulations,
\foamgrid objects are allowed to grow and shrink, i.e., elements can be added and removed from the grid at runtime.
Finite element and finite volume data is kept during such grid changes, using an approach much like the one
used for grid adaptivity.  Of all \dune grids, \foamgrid is currently the only one to support this feature.
The programmer interface combines ideas from the \cpp{GridFactory}
class to insert new vertices and elements, with the adaptivity interface to allow to keep data across
steps of grid growth.

Growth and shrinkage of grids is a two-step process.  First, new elements and vertices are handed to the \cpp{FoamGrid}
object.  These are not inserted directly; rather, they are queued for eventual insertion.  In addition,
individual elements can be marked for removal.  Once all desired elements and removal marks are known to the grid,
the actual grid modification takes place in a second step.

Queuing elements for insertion and removal is controlled by three methods.  The first,
\begin{c++}
std::size_t insertVertex(const FieldVector<double,dimworld>& x);
\end{c++}
queues a new vertex with coordinates \cpp{x} for insertion.  The return value of the method is an index
that can be used to refer to this vertex when inserting new elements.  The index remains fixed
until all queued elements are actually inserted in the grid (by the \cpp{grow} method), but may change during
the execution of that method.  Inserting elements is done using
\begin{c++}
void insertElement(const GeometryType& type,
                   const std::vector<unsigned int>& vertices);
\end{c++}
which mimics the corresponding method from the \cpp{GridFactory} class. The argument \cpp{type} has to be
a simplex type, because (currently) \foamgrid supports only simplex elements.  The array \cpp{vertices} must
contain the indices of the vertices of the new element to be inserted. These can be either indices of existing
vertices, or new indices obtained as the return values of the \cpp{insertVertex} method.

Analogously a new element with a parametrization can be inserted by calling
\begin{c++}
void insertElement(const GeometryType& type,
                   const std::vector<unsigned int>& vertices,
                   const std::shared_ptr<VirtualFunction<
                                             FieldVector<ctype,dim>,
                                             FieldVector<ctype,dimworld>
                                        > >& elementParametrization);
\end{c++}

Finally, the method
\begin{c++}
void removeElement(const Codim<0>::Entity& element);
\end{c++}
marks the given element for removal.

Once all desired elements are queued for insertion or removal, the actual grid modification takes place
in a second step. The grid is modified using the method
\begin{c++}
bool elementsInserted = grid->grow(); // true if at least one element was inserted
\end{c++}
While element removal is guaranteed, queuing elements does not assure that the element will be inserted. New elements are restricted
by the fact that \dune grids are hierarchic objects. The vertices given by the user to form an element are always leaf vertices but may be contained in different hierarchic levels. However, elements can only be constituted by vertices of the same level. Therefore, new elements in
\foamgrid are always inserted on the lowest possible level substituting the given vertex by its
hierarchic descendants or ancestors. Note that it is not generally guaranteed that relatives of
the given vertices on the same level can be found. In that case, the element will not be inserted. The method \cpp{grow} will return \cpp{true} if it was possible to insert at least one
element.

After the call to \cpp{grow}, it is possible to check whether a given element has been created by the last call to the \cpp{grow} method:
\begin{c++}
bool isNew = element.isNew();    // true if element was created by last growth step
\end{c++}
which is a method of the interface class \cpp{Entity<0>}, i.e. elements.
Observe that this is the same method that returns whether an element has been created by grid refinement.
Hence its semantics depends on whether it is queried after a call to \cpp{grow} or after a call to \cpp{adapt}.
Using this method is helpful, e.g., when setting initial values and/or boundary conditions for newly created elements.

The growth is completed with the call
\begin{c++}
grid->postGrow();
\end{c++}
which removes all \cpp{isNew} markers.

Summing up, growing or shrinking the grid while keeping grid data consists of
the following steps.  Note the relationship to grid adaptivity with data transfer.
\begin{enumerate}
 \item Mark elements for removal; queue new vertices and elements for insertion.
 \item Transfer all simulation data attached to the grid into an associative container indexed by the entity ids described in Section~\ref{sec:attaching_data}.
 \item Call \cpp{grow()}.
 \item Resize data array; copy data from the associative container into the array.
 \item Set initial data at newly created elements and vertices and boundary conditions at newly formed boundaries. Note that element removal always creates new boundaries.
 \item Finalize by calling \cpp{postGrow()}.
\end{enumerate}

While grid growth itself is straightforward, it is difficult to use in combination with adaptive refinement.
Grid refinement in \dune leads to hierarchical grids, which are forests of refinement trees.  Not every element
of such a forest can be added or removed without violating certain consistency conditions.  In one-dimensional
grids, there are relatively few problems, and grid growth and refinement can be used together to good effect. For two-dimensional grids we have tried to be as
general as possible, but there are limits.

There are obstacles both to the removal of elements and to the insertion of new ones, if grid refinement is involved.
First of all, only leaf grid elements can be
removed.  There is no conceptual problem with element removal if the element has no father.  If it does have a
father, however, removing only a subset of its $2^d$ sons is a violation of the \dune grid interface
specification, which mandates that the sons of an element must (logically) cover their father~\cite[Def.\,11.1]{dunegridpaperI:08}.
Deliberately allowing this violation nevertheless appears to be the only viable solution, as all other
possibilities amount to only allowing the removal of large groups of elements, which is rarely desired.%
\footnote{Another \dune module which violates this assumption to good effect is \dunemodule{dune-subgrid},
  see~\citep{graeser_sander:2009}.}

Inserting new elements into a refinement forest of elements is even more problematic, because the new element
needs to be assigned a level number in the hierarchy.  Ideally, this level would always be zero, because if the
element had a larger level number it would be expected to have a father.  \foamgrid does violate this assumption
if necessary, which does not lead to problems in practice, unless multigrid-type algorithms are used.
On the other hand, the element level must be the same as the levels of all of its vertices.  If the vertices
are new, their levels can be freely chosen.  However, grid growth almost always involves vertices that already
exist in the grid.  When trying to insert elements with vertices having different level numbers, \foamgrid
currently tries to replace existing vertices by their father vertices, to obtain a set of $d+1$ vertices
on the same level (which then determines the element level).

\section{Numerical examples}
\label{sec:numerical_examples}

We close the article with three numerical examples.  These show how seemingly challenging algorithms can be implemented
with ease using the \foamgrid grid manager.

\subsection{Unsaturated Darcy flow in a discrete fracture network}
\label{sec:example_fracture_network_flow}

For our first example we simulate unsaturated Darcy flow through a network of two-dimensional fractures embedded
into $\mathbb{R}^3$.  We only consider the flow in the network itself, but \foamgrid can be easily coupled to higher-dimensional
background grids using the \dunemodule{dune-grid-glue} module~\citep{bastian_buse_sander:2010,engwer_muething:accepted}.

\begin{figure}
 \begin{center}
  \includegraphics[width=0.5\textwidth]{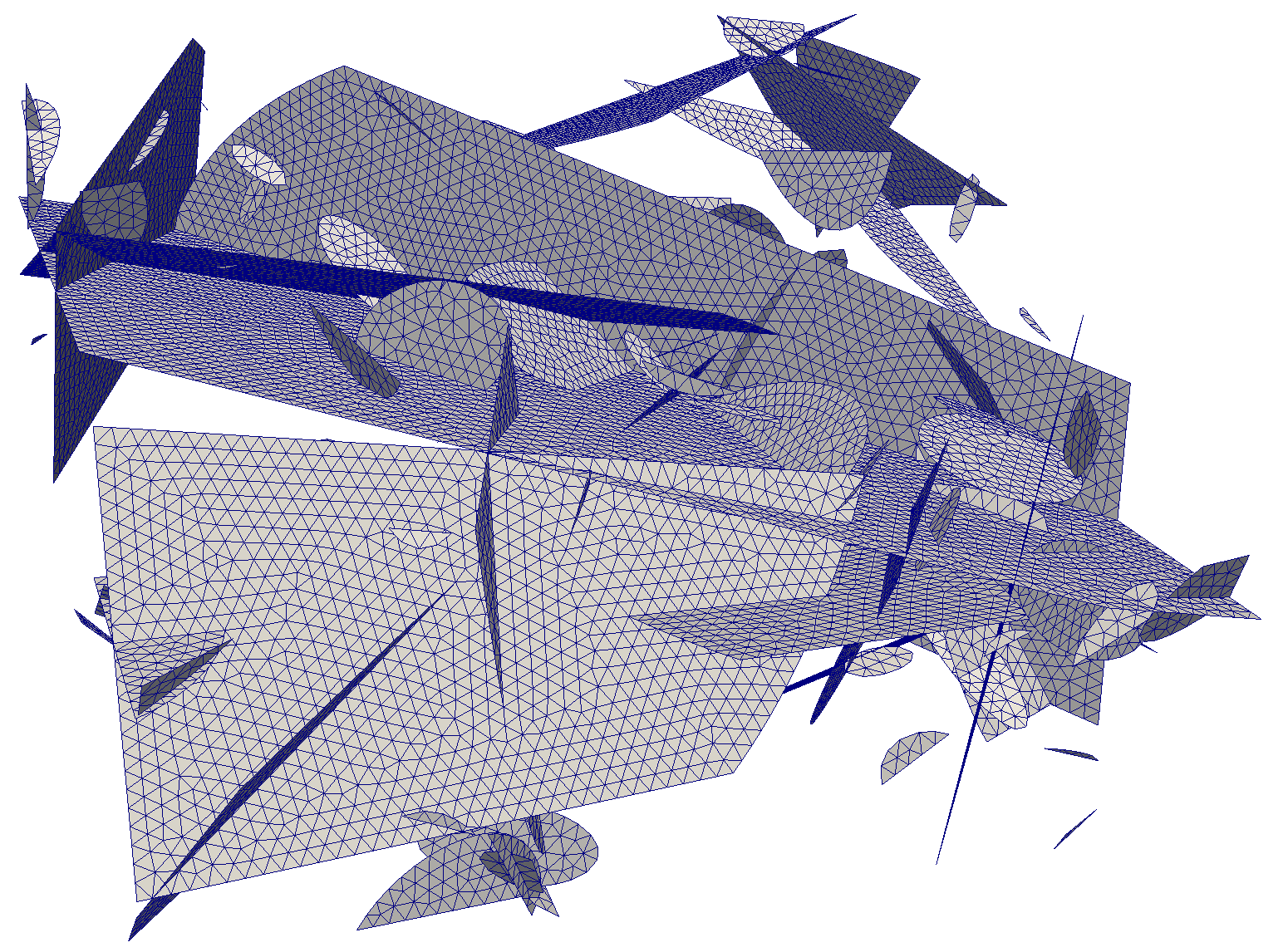}
 \end{center}
 \caption{Grid for a discrete fracture network, courtesy of Patrick Laug.  The network and grid were generated
   using the software described in~\citep{borouchaki_laug_george:2000}.
 }
 \label{fig:fracture_network_grid}
\end{figure}

Let the computational domain $\Omega$ be the union of a finite number of closed, bounded
hypersurfaces in $\mathbb{R}^3$.
We use the Richards equation to model the flow in $\Omega$~\citep{bear:1988}.  That is, we suppose that the state of the
system in a time interval $[0,T]$ can be described by single scalar pressure field
\begin{equation*}
 p : \Omega \times [0,T] \to \mathbb{R}.
\end{equation*}
From this pressure field, the fluid saturation $\theta$ and permeability $\operatorname{kr}$ can be computed using the
Brooks--Corey and Burdine parameter functions
\begin{equation*}
\theta(p)
=
\begin{cases}
\theta_m+(\theta_M-\theta_m)\left(\frac{p}{p_b}\right)^{-{{\lambda}}} & \mbox{ for }\; p\leq p_b\\
\theta_M & \mbox{ for }\; p\geq p_b,
\end{cases}
\qquad\qquad
\operatorname{kr}(\theta) = \left(\frac{\theta-\theta_m}{\theta_M-\theta_m}\right)^{3+\frac{2}{\lambda}},
\end{equation*}
where $\theta_m$, $\theta_M$, $p_b$, and $\lambda$ are scalar parameters.
The saturation $\theta$ satisfies the Richards equation
\begin{equation}
\label{eq:richards_equation}
\frac{\partial}{\partial t}\,\theta(p)+
\operatorname{div} {\,\mathbf v}(x,p)=0,
\qquad
{\mathbf v}(x,p)= -K(x)\operatorname{kr} (\theta(p))\nabla (p - \varrho g z).
\end{equation}
For simplicity we suppose that the flow is purely driven by the boundary conditions, and omit the gravity term $\varrho g z$.

Equation~\eqref{eq:richards_equation} is a quasilinear equation in the pressure $p$.  In \citep{alt_luckhaus:1983} (see also \citep{berninger_kornhuber_sander:2011})
it was shown how the Kirchhoff transformation can be used to transform it to a semilinear equation for a generalized
pressure
\begin{equation*}
u : \Omega \times [0,T] \to \mathbb{R},
\qquad
u(x,t) = u(p(x,t)) \coloneqq \int_0^p\operatorname{kr}(\theta(q(x,t)))\,dq.
\end{equation*}

We discretize this equation in time using an implicit Euler method, and in space using first-order Lagrangian finite elements.
The resulting weak discrete spatial problem can be written as a minimization problem for a strictly convex functional.
At each time step, we determine the minimizer of this functional by a monotone multigrid method \citep{berninger_kornhuber_sander:2011}.

\bigskip

For the implementation we used the code used for the numerical examples in~\citep{berninger_kornhuber_sander:2011}.
Since vertex-based finite elements were used for the discretization, no changes to the numerical algorithm were needed
to also apply it to network grids.
Originally, the code used the \dune libraries and the \uggrid grid manager for unstructured grids.
Even though the code for~\citep{berninger_kornhuber_sander:2011} was not written with network flow problems
in mind at all, it could nevertheless be reused as is, after only a handful of trivial bugfixes.  The only changes
necessary were replacing the line
\begin{c++}
typedef UGGrid<2> GridType;
\end{c++}
by
\begin{c++}
typedef FoamGrid<2,3> GridType;
\end{c++}
and adjusting the boundary data specification.  This once more proves the point that using the \dune grid interface
gives great flexibility, and allows code to do more things than planned by the original authors.

We present an example simulation using an artificially created network grid.  The grid, which can be seen in
Figure~\ref{fig:fracture_network_grid}, was created by Patrick Laug using the fracture network grid generator
described in~\citep{borouchaki_laug_george:2000}.  The grid spans the volume $[-6.5,6.5]^2 \times [-2.165, 2.165]$
(length and pressure are given in meters).  We assume
the network to be filled with a sand-like material with parameters
$\theta_m = 0.0458$, $\theta_M = 1$,
$K = 6.54 \cdot 10^{-5}\,\text{m}/\text{s}$,
bubbling pressure $p_b = 0.0726$\,m, and $\lambda = 0.694$.

\begin{figure}
 \begin{center}
  \includegraphics[width=0.32\textwidth]{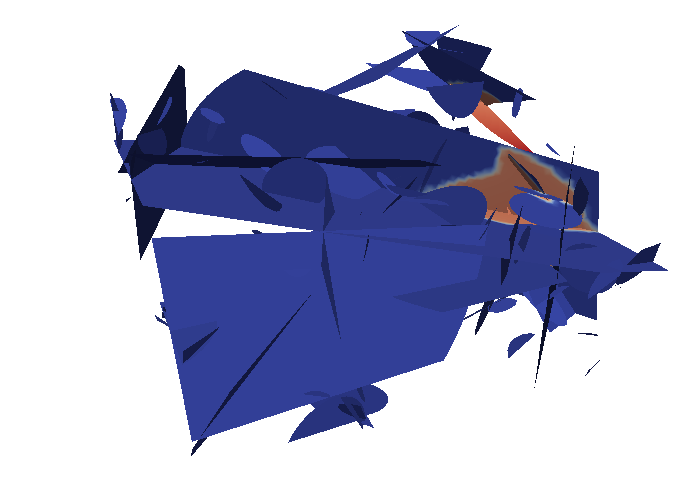}
  \includegraphics[width=0.32\textwidth]{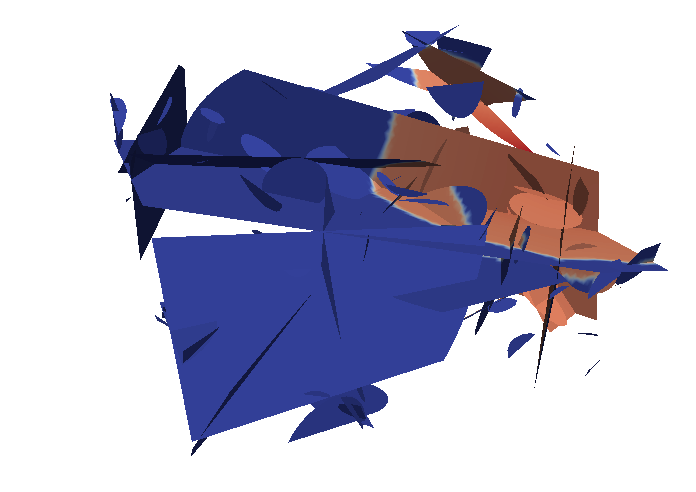}
  \includegraphics[width=0.32\textwidth]{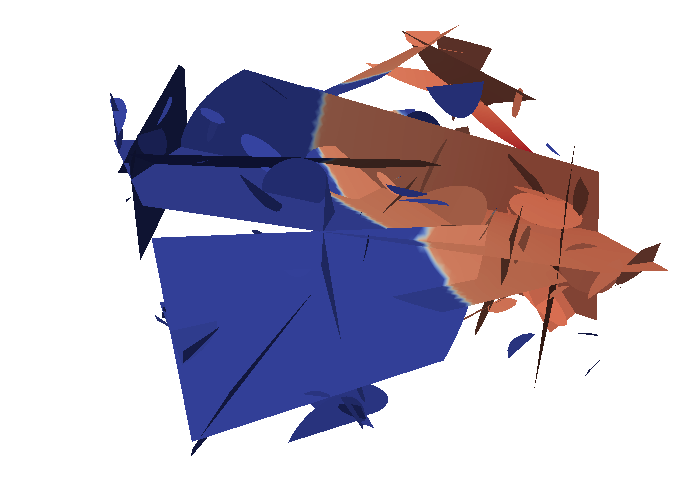}

  \includegraphics[width=0.32\textwidth]{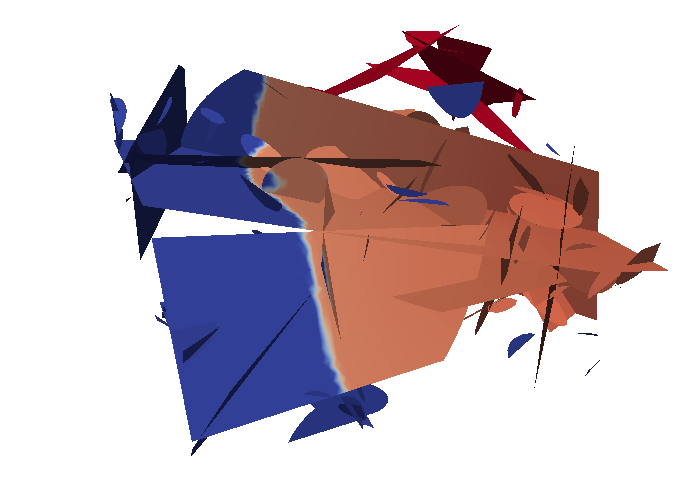}
  \includegraphics[width=0.32\textwidth]{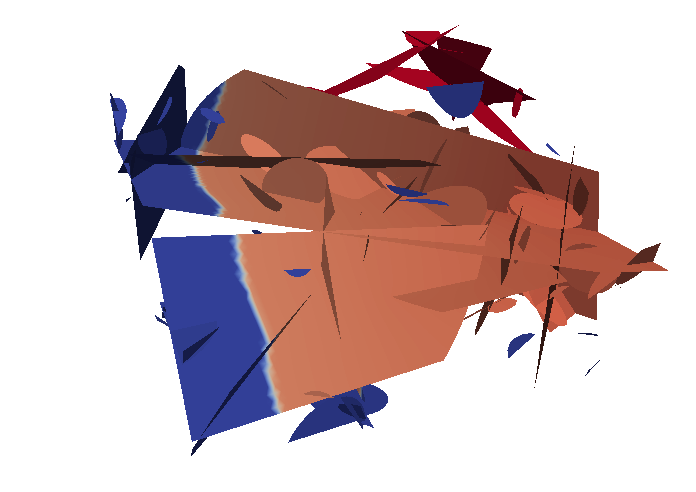}
  \includegraphics[width=0.32\textwidth]{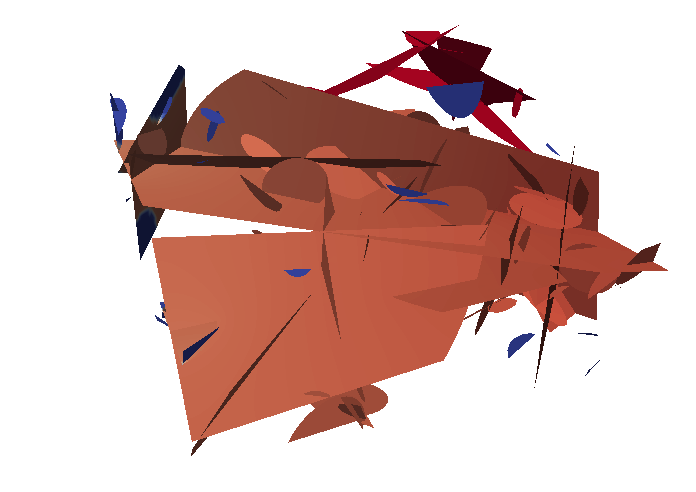}
 \end{center}
 \caption{Unsaturated Darcy flow in a fracture network.  The color visualizes the physical pressure~$p$.}
 \label{fig:fracture_network_flow}
\end{figure}

We assume the network to be initially devoid of water, setting $p=-10$\,m.
Then, water is injected in a unit circle centered at the point $(0,6.5,0)$ on the boundary $\partial\Omega \cap \{x_1 = 6.5\}$
with a constant pressure of $p=3$\,m; no-flow boundary conditions are imposed
at the remaining boundary.  Figure~\ref{fig:fracture_network_flow} shows several steps in the evolution
of the physical pressure $p$.  As expected, one can see the fluid entering, and slowly filling almost
the entire network.  At the end, a steady state is reached, and the pressure is constant in each
connected component of the domain.

\subsection{Transport of a therapeutic agent in the microvasculature}
\label{sec:bloodflow}

\begin{figure}
 \begin{center}
  \includegraphics[width=0.49\textwidth]{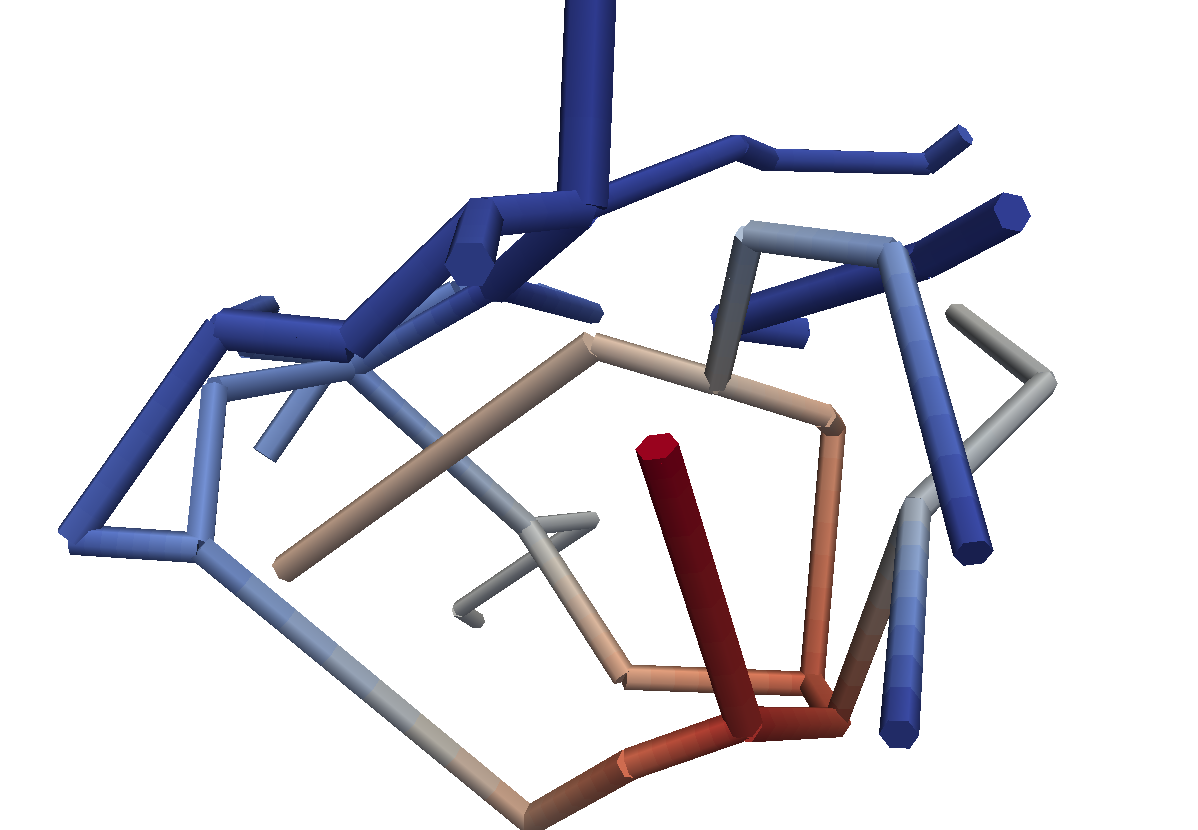}
  \includegraphics[width=0.49\textwidth]{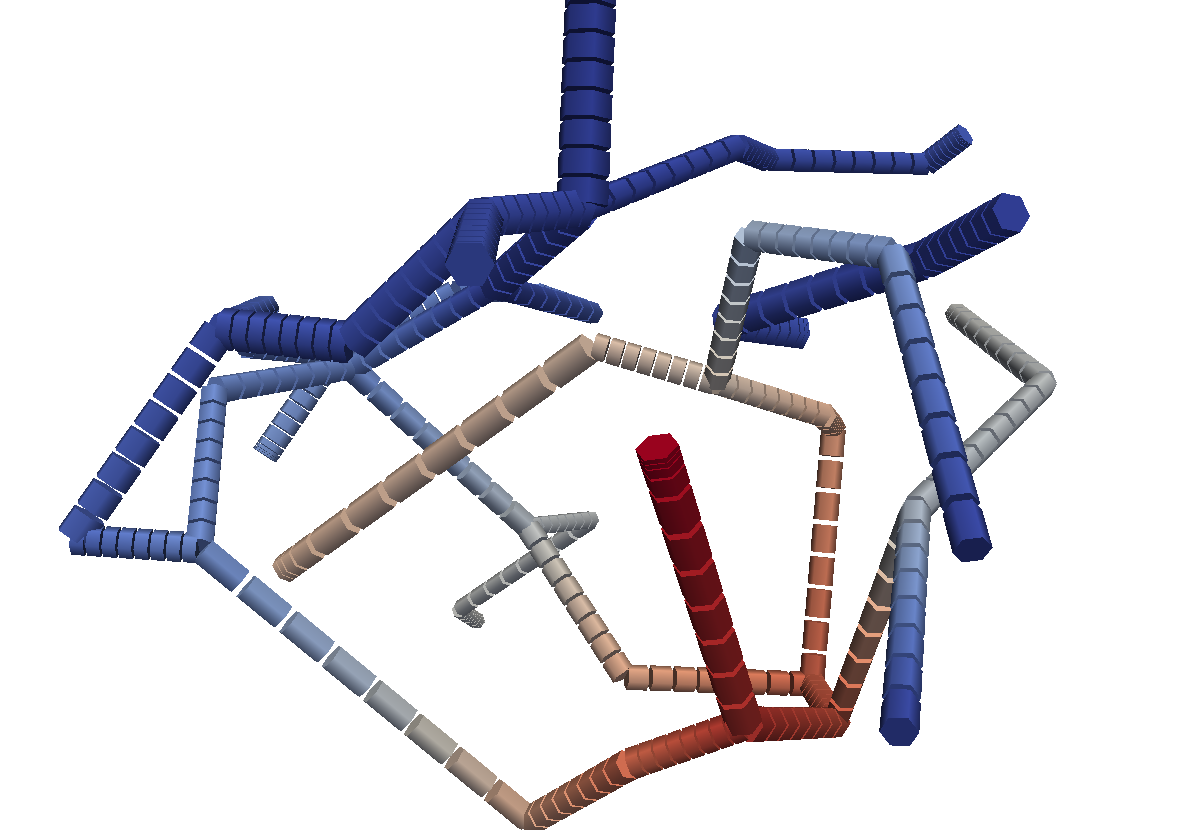}\vspace{0.05\textwidth}
 \end{center}
  \caption{Single-phase two-component flow in a blood vessel network. One-dimensional elements are rendered as tubes.
  The color visualizes the fluid pressure.  Left: stationary pressure $p$; right: initial element sizes.}
 \label{fig:bloodflow1}
\end{figure}

The next example demonstrates local grid adaptivity, and the treatment of network bifurcations.
We model the propagation of a therapeutic agent for cancer therapy in
a network of small blood vessels. To reduce the computational effort, the network of three-dimensional vessels is
reduced to a one-dimensional network $\Omega$ embedded in a three-dimensional tissue domain $\mathcal{T}$.

We first discuss the one-dimensional model for flow in a blood vessel segment, disregarding any bifurcations.
Starting from a three-dimensional blood vessel domain $\Xi$, we describe blood in the microvasculature
as an incompressible Newtonian fluid with viscosity $\mu$ and density $\varrho$ governed by the Stokes equation.
Further, we assume an axially symmetric blood vessel segment with constant radius $R$, cross-section area $A$, the parametrized tangent on the vessel centerline $\lambda(s)$, $\lambda : \mathbb{R} \rightarrow \mathbb{R}^3$, $s \mapsto \lambda(s)$, and a rigid vessel wall. Then, with the assumption of constant pressure $p$ in a cross-section and
negligible radial velocities $v_r$ the Stokes equations can be integrated over a vessel segment yielding the one-dimensional equation
\begin{equation}\label{eq:1dstokes}
\left.
\begin{aligned}
  \frac{A}{\varrho} \frac{\partial p}{\partial s}\lambda + 2\pi \frac{\mu}{\varrho} (2+\gamma) \bar{v} \lambda &= 0\\
  \frac{\partial}{\partial s} \left( \frac{\pi R^4}{2 \mu (2 + \gamma)} \frac{\partial p}{\partial s} \right) &= 0
\end{aligned}  \quad \right\} \quad \text{in }\Omega.
\end{equation}
The parameter $\gamma$ shapes a power-type axial velocity profile with mean velocity $\bar{v}$,
yielding a quadratic velocity profile for $\gamma = 2$ and flatter profiles for $\gamma > 2$.
For a detailed derivation in a more general setting, see the reduction of the Navier--Stokes equations to one-dimensional equations in \citep{quarteroni_formaggia:03}.
Equation~\eqref{eq:1dstokes} is a simplified stationary version of the one-dimensional blood flow equations described by \citet{quarteroni_formaggia:03}, neglecting vessel wall displacement and intertial forces.

Small blood vessels are exchanging mass with the embedding tissue through the vessel wall.
The fluid exchange can be modeled by Starling's law, and results in an additional source term. With this modification, \eqref{eq:1dstokes} becomes
\begin{equation}\label{eq:1dstokes_source}
\left.
\begin{aligned}
  \frac{A}{\varrho} \frac{\partial p}{\partial s}\lambda + 2\pi \frac{\mu}{\varrho} (2+\gamma) \bar{v} \lambda &= 0\\
  \frac{\partial}{\partial s} \left( \frac{\pi R^4}{2 \mu (2 + \gamma)} \frac{\partial p}{\partial s} \right) - 2\pi R L_p (p - \bar{p}_i) &= 0
\end{aligned} \quad \right\} \quad \text{in }\Omega,
\end{equation}
where $L_p$ is the empirical filtration coefficient dependent on, e.g., the intrinsic permeability and thickness of the membrane, and the
viscosity of the interstitial fluid. The source term further depends on the pressure in the surrounding tissue $\bar{p}_i$. For the sake of simplicity, this tissue pressure is subsequently assumed constant. Equation~\eqref{eq:1dstokes_source} was also used by \citet{cattaneo2014computational} in a finite element setting to model
coupled vessel-tissue flow processes in a tumor tissue.

The transport of a therapeutic agent is modeled by an advection--diffusion equation using the velocity field
calculated by equation~\eqref{eq:1dstokes_source}.
Similar to the reduction of the Stokes equations we can reduce the three-dimensional advection--diffusion equation by integration over a
vessel segment assuming a constant concentration $c = x \varrho$ on a given cross-section with area $A = \pi R^2$ \citep{dangelo:2007,cattaneo2014drug}.
The transport over the vessel wall can be described by the Kedem--Katchalsky equation \citep{kedem:1958}, yielding
\begin{equation}\label{eq:1dtransport_source}
  \frac{\partial (Ac)}{\partial t} + \bar{v} \frac{\partial (Ac)}{\partial s} - D_e \frac{\partial^2 (Ac)}{\partial s^2}  - 2\pi R \left[ L_c (c - \bar{c}_i) +  L_p (p - \bar{p}_i) (1 - \sigma_c) c \right] = 0
  \quad \text{in }\Omega.
\end{equation}
The last term in \eqref{eq:1dtransport_source}, accounting for transport across the vessel wall, consists of an advective and a diffusive part where advection is reduced by the reflection coefficient $\sigma_c \in [0,1]$ for larger molecules. Again, we assume the mean tissue concentration $\bar{c}_i$ to be constant.

\bigskip

To model a network of such segments we split the blood vessel network $\Omega$ at junctions into pieces yielding a
set of vessel segments $\Omega_i$ each governed by equations \eqref{eq:1dstokes_source} and \eqref{eq:1dtransport_source}. At each junction we require continuity of pressure
\begin{equation*}
 p = p_1 = ... = p_j.
\end{equation*}
Together with these coupling conditions, and boundary conditions on $\partial \Omega$, Equation~\eqref{eq:1dstokes_source}
has a unique solution, which is the stationary pressure field~$p$.

For the transport at the junctions, we require continuity of concentration
\begin{equation*}
 c = c_1 = ... = c_j,
\end{equation*}
and, for $Q_i$, the volume flux leaving segment $\Omega_i$ at the junction, we require mass conservation
\begin{equation*}
 \sum_{i=1}^j Q_i = 0.
\end{equation*}

We discretize equations \eqref{eq:1dstokes_source} and \eqref{eq:1dtransport_source} in space with a standard finite volume scheme and piecewise linear one-dimensional grid elements. This demands balancing fluxes over the edges of the elements. Assume that we have calculated all element transmissibilities
\begin{equation}
  t_i \coloneqq \frac{\pi R_i^4}{2\mu(2+\gamma)}.
\end{equation}
Then, the volume flux $Q_{ij}$ from element $i$ to a neighboring element $j$ can be calculated as
\begin{equation}
  Q_{ij} = t_{ij}(p_i-p_j) = \frac{t_i t_j}{\sum_{k=0}^{N} t_k}(p_i-p_j).
\end{equation}
One can see that the transmissibility at branching points is dependent on the transmissibility of all $N$ neighboring elements.
Assuming that all element transmissibilities $t_i$ reside in an array \cpp{transmissibility}, the calculation of the two-point transmissibilities $t_{ij}$
could look as follows using \foamgrid. Note that only a single loop over the grid entities is necessary to calculate the transmissibilities.
\begin{c++}
// for each element with index eIdx
std::vector<double> tSums(e.subEntities(/*codim=*/ 1), 0.0);
std::vector<double> tij;
std::vector<std::size_t> neighborFacetMap;

// loop over all intersections of this element
for(const auto& intersection : intersections(foamGridLeafView, element))
{
  if(intersection.neighbor())
  {
    std::size_t nIdx = foamGridLeafView.indexSet().index(intersection.outside());
    tij.push_back(transmissibility[eIdx]*transmissibility[nIdx]);
    neighborFacetMap.push_back(intersection.indexInInside());
    tSums[intersection.indexInInside()] += transmissibility[eIdx];
  }
  if(intersection.boundary())
    // boundary treatment ...
}

// compute the two-point transmissibilities
for (std::size_t i = 0; i < tij.size(); i++)
  transmissibilitiesIJ[i] /= tSums[neighborFacetMap[i]];
\end{c++}

This code works well using the grid interface of \dunemodule{dune-grid-2.4}, even though that interface does not
have any provisions for network grids at all (Section~\ref{sec:intersections}).  This is because the grid is
one-dimensional.  In this case, each intersection always covers entire element facets (i.e., vertices).
Therefore, the \cpp{indexInInside} method can be used to group intersections.

\begin{figure}
 \begin{center}
 \begin{subfigure}{0.45\textwidth}
  \includegraphics[width=\textwidth]{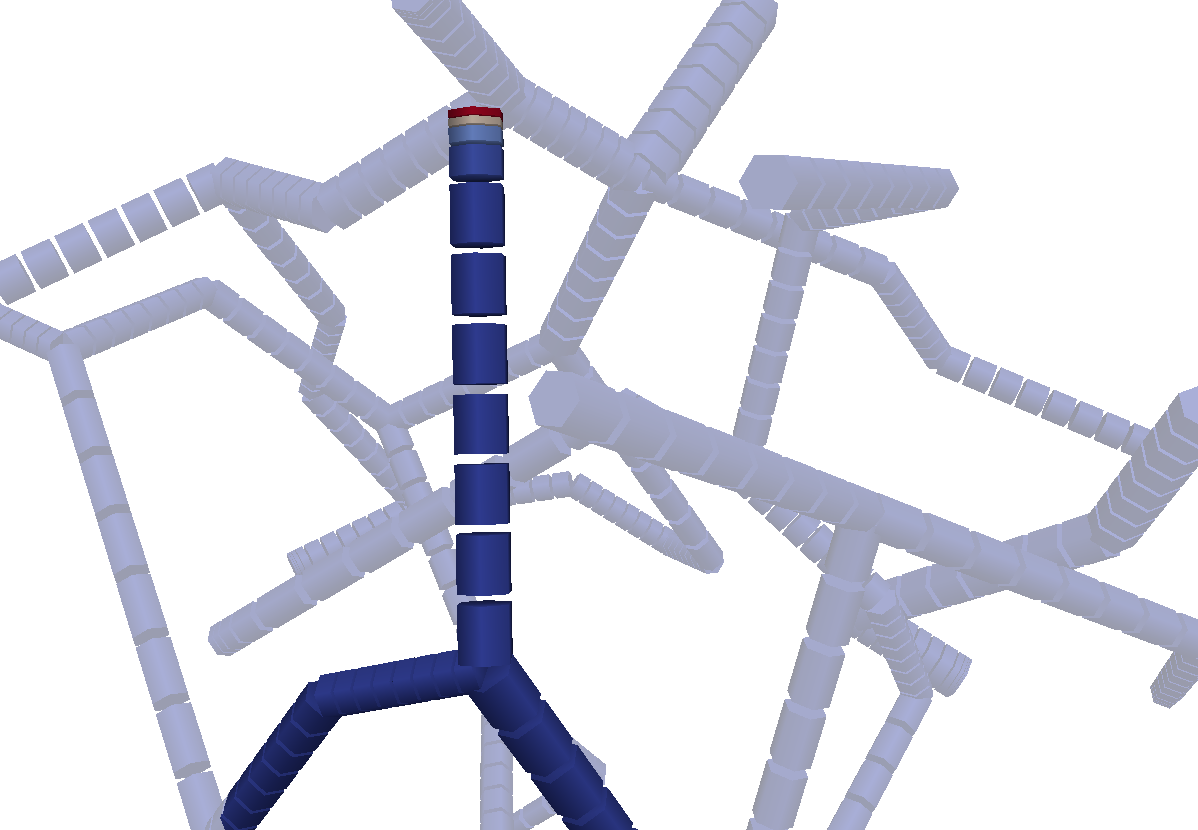}
  \caption{$t=3$\,s}
 \end{subfigure}
 \hspace{0.05\textwidth}
 \begin{subfigure}{0.45\textwidth}
  \includegraphics[width=\textwidth]{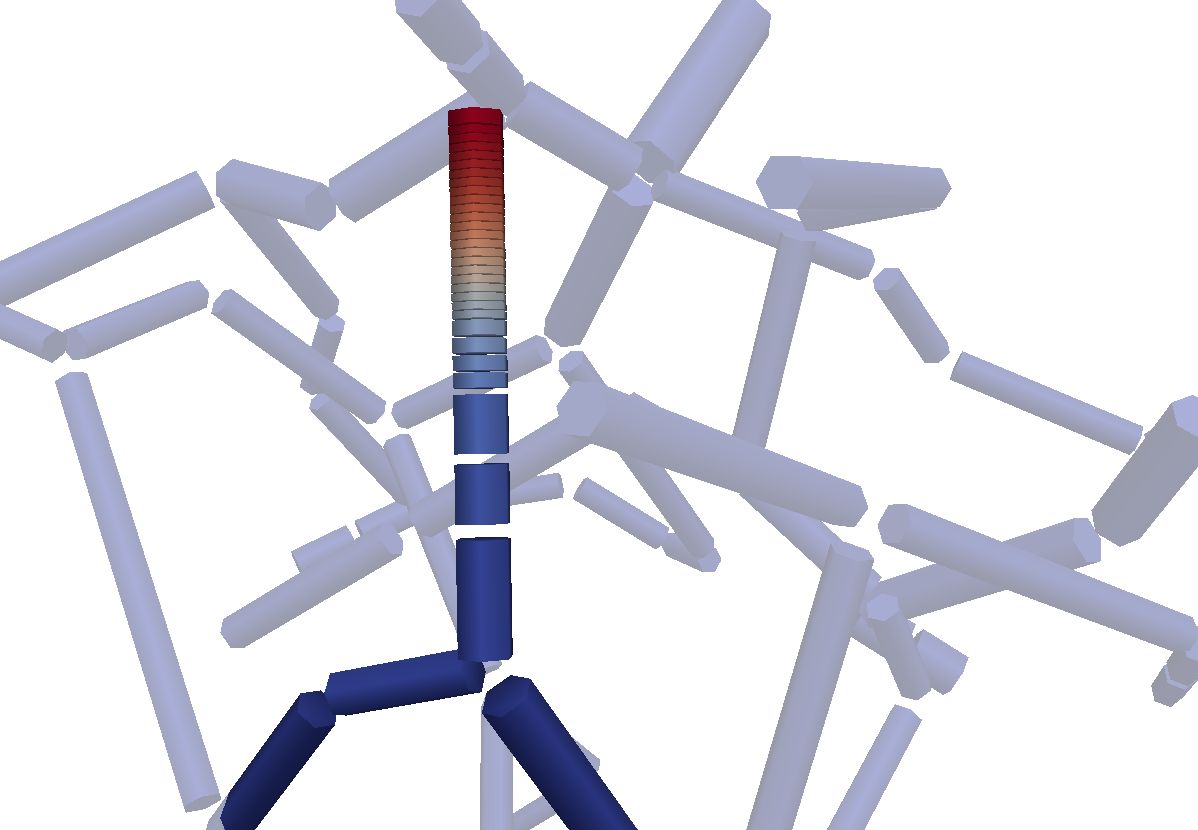}
  \caption{$t=56$\,s}
 \end{subfigure}
 \vspace{0.04\textwidth}

 \begin{subfigure}{0.45\textwidth}
  \includegraphics[width=\textwidth]{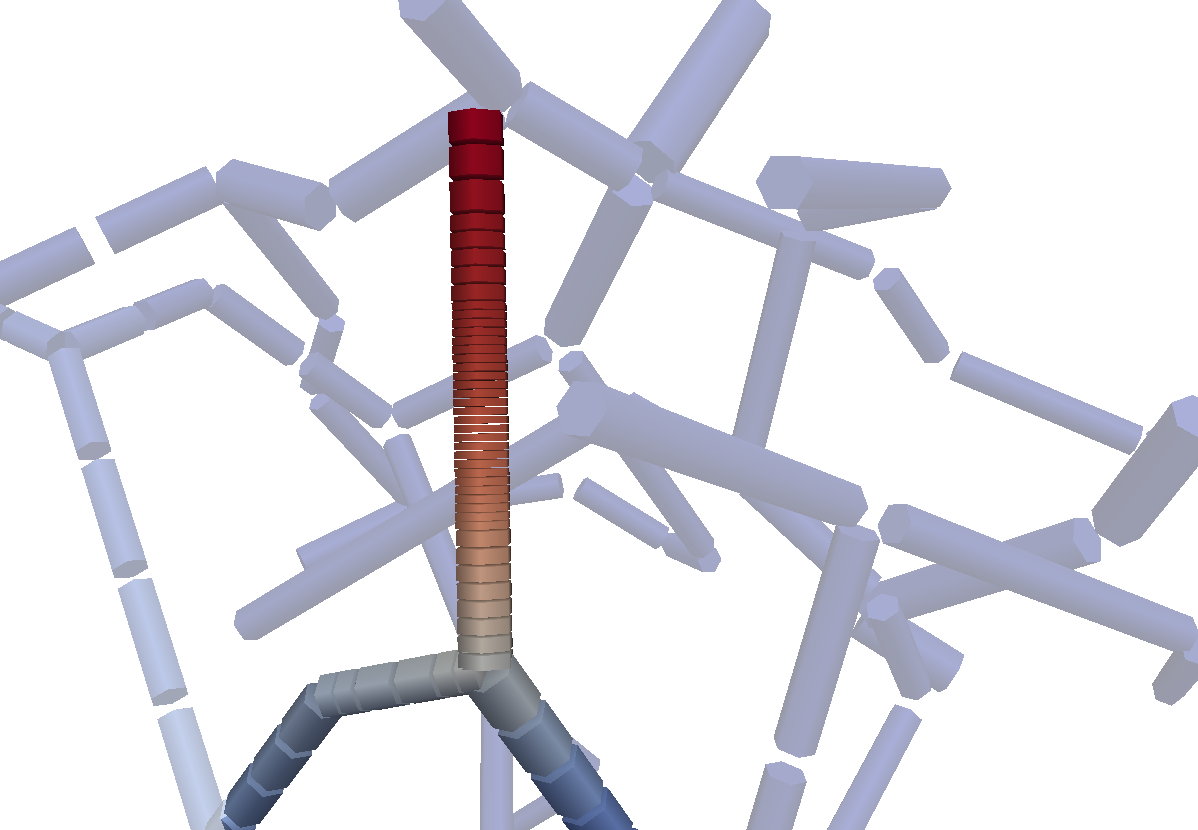}
  \caption{$t=201$\,s}
 \end{subfigure}
 \hspace{0.05\textwidth}
 \begin{subfigure}{0.45\textwidth}
  \includegraphics[width=\textwidth]{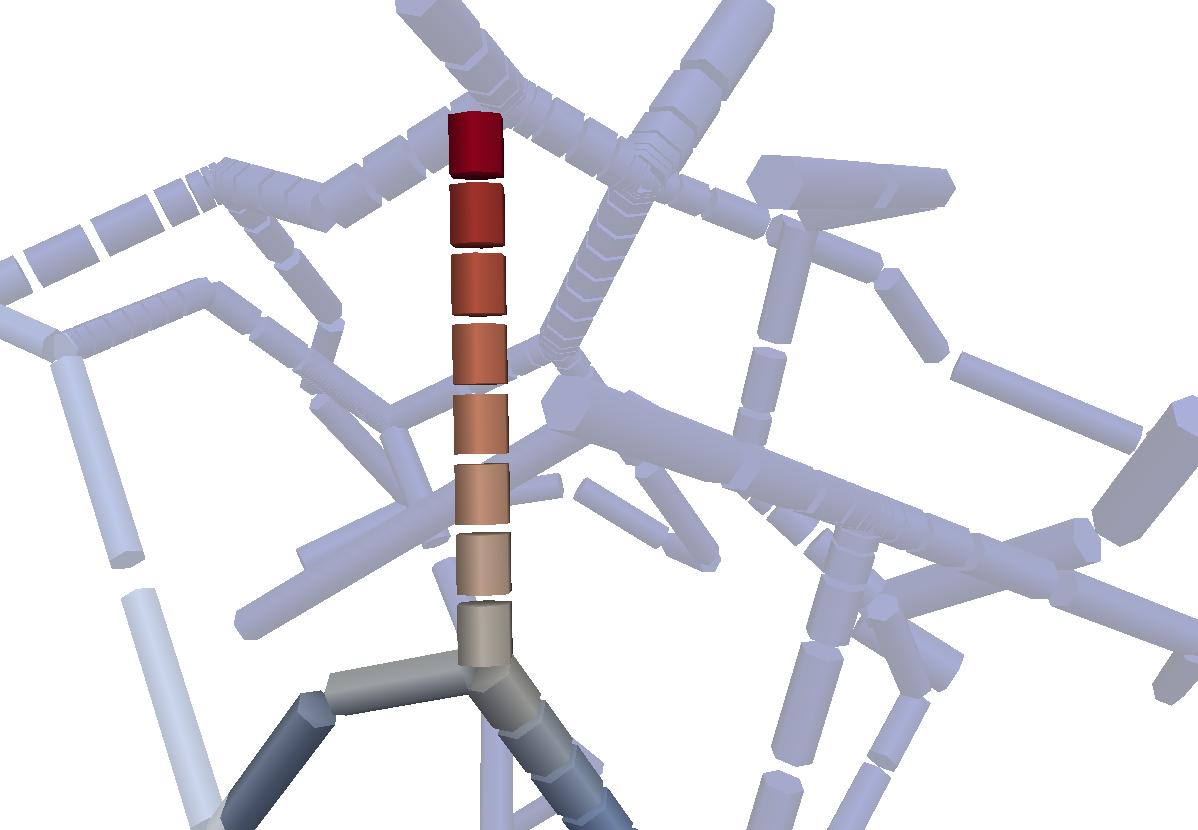}
  \caption{$t=3\,000$\,s}
 \end{subfigure}
 \end{center}
 \caption{Single-phase two-component flow in a blood vessel network. One-dimensional elements are rendered as tubes. The gaps merely exist to better visualize the adaptive grid. The images show the mole fraction $x$ at different simulation times.
 Note how a concentration wave enters the network from the top, and is tracked by a zone of high grid
 resolution, even as it goes through a bifurcation point.}
 \label{fig:bloodflow2}
\end{figure}

\bigskip

To reduce numerical diffusion induced by the implicit Euler scheme, the grid can be refined adaptively around the concentration front.
Initially, the grid is refined around inflow boundaries, and during the simulation the grid is adapted by a
local gradient-based concentration indicator as described in \citep{wolff:2013}. This
indicator marks an element $i$ for refinement if
\begin{equation*}
\frac{\max_i(\Delta c_{ij} ) - \Delta c_\text{min}}{\Delta c_\text{max} - \Delta c_\text{min}} \geq \epsilon_r,
\end{equation*}
and for coarsening if
\begin{equation*}
\frac{\max_i(\Delta c_{ij} ) - \Delta c_\text{min}}{\Delta c_\text{max} - \Delta c_\text{min}} < \epsilon_c,
\end{equation*}
where $\Delta c_{ij}$ denotes the concentration difference between element $i$ and a neighboring element $j$.
The parameters $\epsilon_r$, $\epsilon_c \in [0,1]$ ($\epsilon_r > \epsilon_c$) are problem dependent.
We achieved robust adaptation behavior in our example for $\epsilon_r = 0.3$ and $\epsilon_c = 0.05$.
The indicator is evaluated before every time step and marks elements for refinement or coarsening using the \cpp{mark} method, see Section~\ref{sec:adaptive_refinement}.
The adaption of the grid is then handled by \foamgrid. Between the \cpp{preAdapt} and \cpp{postAdapt} steps, we have to transfer data, i.e.,
primary variables and spatial parameters from the old grid to the new adapted grid.
As can be seen in Figure~\ref{fig:bloodflow2}, the refinement scheme works well even around network bifurcations.

We simulate a network of capillaries in rat brain tissue scanned by~\citet{motti:1986} and reconstructed as segment network with three-dimensional geometrical information by~\citet{secomb:2000}. The network data comprises three-dimensional location of vessel segments, inflow and outflow
boundary markers, vessel segment radii, and velocity estimates for each vessel segment.
The domain has a bounding box of $150\,\mu\text{m} \times 160\,\mu\text{m} \times 140\,\mu$m. The given vessel radii vary between $2\,\mu$m and $4.5\,\mu$m. The estimated velocities are in a range of $0.5\,\frac{\text{mm}}{\text{s}}$ to $7.5\frac{\text{mm}}{\text{s}}$. We choose the blood viscosity $\mu = 3.0 Pa\cdot s$, the filtration coefficient $L_p = 3.33\cdot 10^{-12} \frac{m}{Pa \cdot s}$, the effective diffusion coefficient over the vessel wall $L_c = 10^{-5} \frac{1}{sm}$, and the diffusion coefficient of the transported agent trail $D_e = 2.93\cdot 10^{-14}\,\frac{1}{\text{s}}$, calculated with the Stokes--Einstein radius.
We assign the following boundary conditions for the flow problem~\eqref{eq:1dstokes_source}
\begin{equation*}
\begin{aligned}
  p &= p_D \quad &&\text{on } \partial\Omega_{\text{outflow}},\\
  \left[ \frac{R^2}{\mu (2 + \gamma)} \frac{\partial p}{\partial z} \right] \cdot n = \bar{v} \cdot n &= v_N \quad &&\text{on } \partial\Omega_{\text{inflow}}.
\end{aligned}
\end{equation*}
The velocities for the inflow segments are set to the estimates provided by~\citet{secomb:2000} along with the grid geometry.
We simulate the arrival of a therapeutic agent by a Dirichlet boundary condition for equation~\eqref{eq:1dtransport_source} on a subset $\partial\Omega_{\text{c}}$ of $\partial\Omega_{\text{inflow}}$, namely, 
\begin{alignat*}{2}
  c &= c_D &\qquad &\text{on } \partial\Omega_{\text{c}}, \\
  c &= 0 &&\text{on } \partial\Omega_{\text{inflow}} \setminus \partial\Omega_{\text{c}}.
\end{alignat*}
Specifically, we enforce a mole fraction of $x_D = 10^{-8}$ at one of the inflow vessel segments with a Dirichlet boundary condition.
At outflow boundaries, we neglect diffusive fluxes. Note that a full upwind scheme is employed for the concentration, so no further boundary condition for the advective fluxes is necessary at outflow boundaries.

Figure~\ref{fig:bloodflow1} shows the resulting stationary pressure field and the initial element size.
In Figure~\ref{fig:bloodflow2}, one can see the resulting mole fraction at various times. Note how the local grid refinement follows the steepest gradients, i.e., the transport front. The resulting mole fractions vary from segment to segment due to different radii. This also automatically ensures a finer grid around vessel bifurcations. The one-dimensional elements are depicted as three-dimensional tubes scaled with their respective radius.

\subsection{Root water uptake and root growth at plant-scale}
\label{sec:root_networks}

In environmental and agricultural research fields, models describing root architectures are used to investigate water uptake and root growth behavior of plants~\citep{Dunbabin2013}. In our final example, a one-dimensional network embedded into $\mathbb{R}^3$ is used to describe such a plant root architecture. We simulate water flow through the root network and root growth.

Plant roots can be described by a tree-like network of pipes which consist of xylem tubes \citep{Tyree2002}. We follow the cohesion–tension theory \citep{Tyree1997}, where the water flow through the root system is governed by the pressure gradient caused by the transpiration rate of the plant above the soil.
We assume only vertical flow and no gravity. This leads to a Darcy's law analogy \citep{Doussan1998}
\begin{equation*}
q_x = -K_x \frac{d p_x}{d z},
\end{equation*}
where $q_x$ is the water flux in the xylem tubes, $p_x$ is the xylem water pressure, and $K_x$ is the axial conductance of one root segment.

Water can enter the roots at any point on the xylem tube surfaces, which leads to a volume source term for the one-dimensional network model.
For simplicity, we model a single membrane only for the entire pathway of water from soil into the roots. With this assumption, and neglecting osmotic processes, radial water flow $q_r$ into one root segment is defined as
\begin{equation*}
q_r = K_r A_r (p_S - p_x),
\end{equation*}
where $K_r$ is the radial conductivity, i.e., the conductivity of series of tissues from root surface to the xylem. The number $A_r \coloneqq 2\pi rl$ is the soil--root interface area. The water pressure $p_S$ at the soil--root interface must be provided. One option is to couple the root system to a Richards equation based soil water flow simulation \citep{Javaux2008}, but for simplicity we simply take $p_S$
as a known value.

The continuity equation leads to
\begin{equation}
	-\operatorname{div}(q_x(p_x)) =  S(p_x)
	\label{root:conti}
\end{equation}
with a solution-dependent source
\begin{equation*}
S(p_x) = K_r A_r (p_S - p_x),
\end{equation*}
where $q_x$ is the only unknown variable.
This modeling approach  neglects the  influence of solutes on water flow, as well as  the capacitive effect of the roots, because the amount of
water stored in roots is generally small compared to transpiration requirements.

Equation~\eqref{root:conti} is discretized in space with standard cell-centered finite volumes, piecewise linear one-dimensional grid elements and implemented using the external \dune discretization module \dumux~\citep{Flemisch2011}, and the \foamgrid grid manager.
Flux calculations over edges and branching points of the root network are implemented just as in our previous example.

\bigskip

So far, we have assumed a root network that does not alter its geometry over time.
Several algorithms were developed to describe root growth (e.g., \citep{Pages2004,Somma1998,Leitner2010b}). These models define root growth
either by a fractal description, or more generically depending on plant specific parameters (root elongation, growth direction, branching density)
and surrounding soil properties (soil moisture, soil strength, temperature, nutrients). For simplicity, we model root growth here as an (almost) completely
random process. New root branches occur at random time steps and with a small gravity effect only, which makes the roots tend to point downwards.
Existing branches grow at the branch tip at random times and without changing directions.

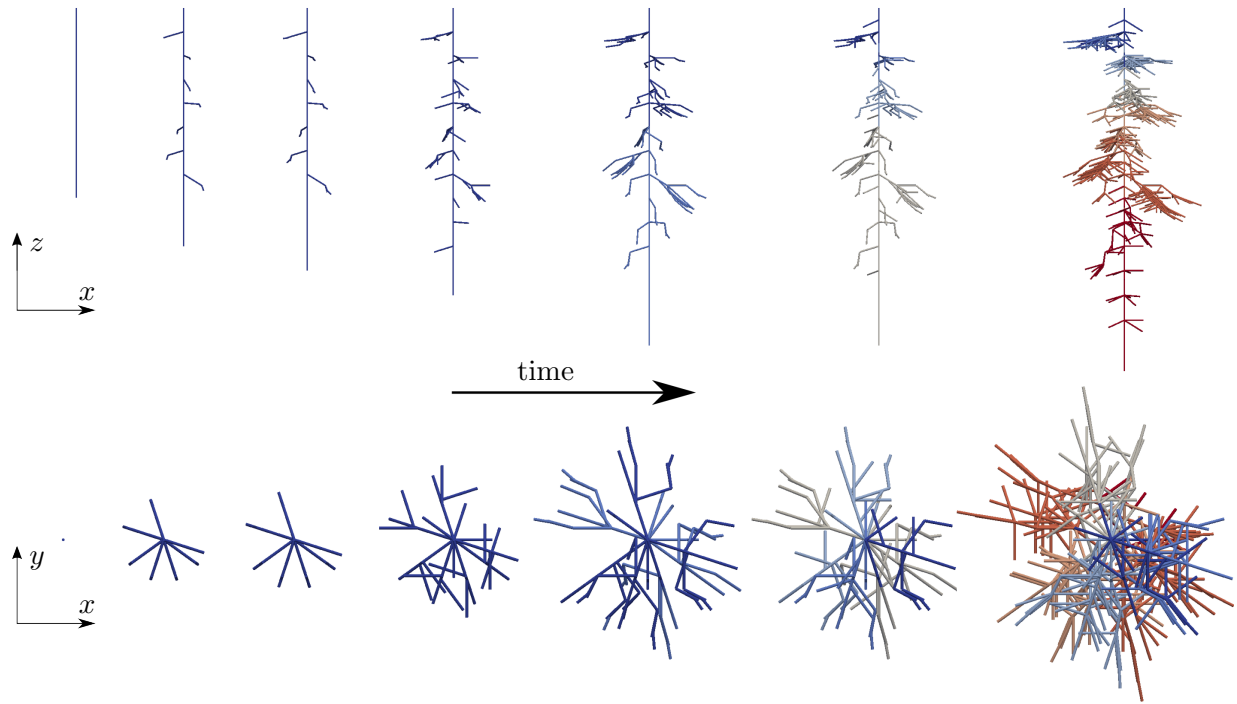
\begin{figure}
	\begin{center}
	\begin{overpic}[width=\textwidth]{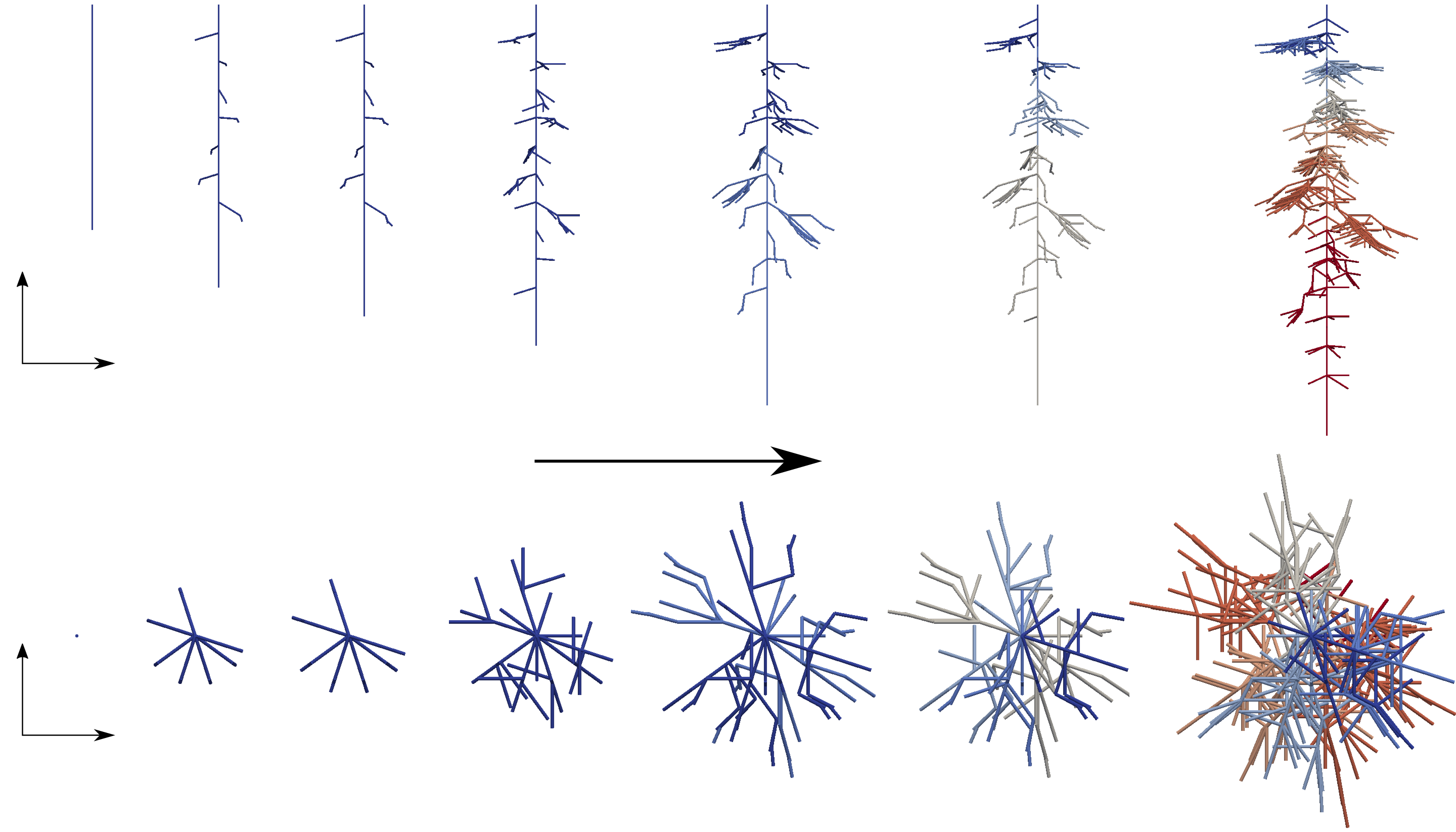}
	 \put(6.5,7.5){$x$}
	 \put(6.5,33){$x$}
	 \put(2.5,11.5){$y$}
	 \put(2.5,37){$z$}
	 \put(42,26.5){time}
	\end{overpic}
	\end{center}
	\caption{Growth of a root system, shown in a lateral (top) and an axial (bottom) view.
	         The color represents the pressure inside the roots.}
	\label{fig:rootGrowth}
\end{figure}

Our example simulation starts with a simple root grid which consist of one vertical root branch discretized with eight elements (root segments)
and no lateral branches (Figure~\ref{fig:rootGrowth}). We choose radial and axial conductivity values from \citep{Doussan1998}.
The surrounding relative soil pressure is set to $p_S = -2.9429 \cdot 10^{-2}$\,Pa, and the Dirichlet boundary value at the root collar is
set to $-1.2 \cdot 10^{6}$\,Pa. Parameters and boundary conditions do not change with time.

In every time step, new root elements are created and either added to an existing lateral branch or to the main branch. An element-based indicator based on simulation time and random factors decides whether a new branch is added at one of the element's vertices.
The \cpp{Indicator} class also computes the coordinates of the newly inserted point that is the second vertex of the new element. The coordinates depend again on simulation time, random factors, and the branch orientation in $\mathbb{R}^3$.
Our growth step calculation in \dumux  using \foamgrid looks as follows.
\begin{c++}
template<class Indicator>
void growGrid(Indicator& indicator, Variables& vars)
{
  // (1) calculate indicator for each element
  indicator.calculateIndicator();

  // (2) insert elements according to the computed indicator
  insertElements(indicator);

  // (3) Put variables in a persistent map
  storeVariables(vars);

  // (4) Grow grid
  grid->grow();

  // (5) Resize and (re-)construction of variables
  vars.resize(foamGridLeafView.size(0));
  reconstructsVariables(vars);

  // (6) delete isNew markers in grid
  grid->postGrow();
}
\end{c++}

The \cpp{vars} container contains all primary variables and spatial parameters defined on the root segments. The \cpp{Indicator} class is a template parameter that can be easily exchanged, to allow different root growth algorithms. In Step~(2), the new elements are inserted.
New root segments must be connected to the old grid.  The implementation of the \cpp{insertElements} method could looks as follows:
 \begin{c++}
template<class Indicator>
void insertElements(const Indicator& indicator)
{
  // iterate over all grid elements (root segments)
  for (const auto& element : elements(foamGridLeafView))
  {
    // find elements that will get a new neighbor
    if (indicator.willGrow(element))
    {
      // get the new elements vertices from the indicator
      std::size_t vIdx0 = grid->insertVertex(indicator.getNewVertexCoordinates(element));
      // get index of the existing element vertex the new element will be connected to
      std::size_t vIdx1 = indicator.getConnectedVertex(element);
      // insert new element with the two vertex indices
      grid->insertElement(element.type(), {vIdx0, vIdx1});
    }
  }
}
 \end{c++}
The primary variables and spatial parameters of our physical problem are stored in Step~(3), before the actual growth step (4). We update the sizes
of the variables and parameter vectors since the total number of degrees of freedom has changed due to the growth step (5). In addition, values for the new elements have to be computed. In our case, new root segments inherit the primary variables and the spatial parameters from its preceding neighbor element. Root tips, in particular new root tips, are always assigned Neumann no-flow boundary conditions.
At the end, the \cpp{postGrow} method is called, which deletes the \cpp{isNew} markers (Step~(6)).

 Figure~\ref{fig:rootGrowth} shows the root network and the pressure distribution inside the roots for several time steps.  The total root water uptake (transpiration demand) of the plant, defined by the Dirichlet boundary condition, does not change during the growing period. Thus, the pressure inside the plant changes since the water uptake of the plant is distributed to more and more roots segments.

\bibliographystyle{abbrvnat}
\bibliography{sander_koch_schroeder_flemisch_foamgrid}
\end{document}